\pdfoutput=1

\documentclass[3p,times,twocolumn]{elsarticle}
\usepackage{epsfig}
\usepackage{graphicx}
\usepackage{amsmath}
\usepackage{amssymb}
\usepackage{dcolumn}
\usepackage{bm}
\usepackage{subfig}
\usepackage{booktabs}
\usepackage{multicol}
\usepackage{multirow}
\usepackage{mathtools}
\usepackage[figuresright]{rotating}
\usepackage{amssymb}
\usepackage{color}
\begin{document}
	%
%	\linenumbers
	\begin{frontmatter}
		% notes:
		% use the corauthref command within \author for corresponding author
		% footnotes;
		% \corauth[Shuaib Ahmad khan]{}
		\title{Optimization of multi-gigabit transceivers for high speed data communication links in HEP Experiments}
		
		\author[label1]{Shuaib Ahmad Khan\corref{cor1}}
		\cortext[cor1]{Corresponding author}
		\ead{shuaib.ahmad.khan@cern.ch}
		\author[label1]{Jubin Mitra} 
		\author[label1]{Tushar Kanti Das}
		\author[label1,label2]{Tapan K. Nayak}
		
		\address[label1]{Variable Energy Cyclotron Centre, Homi Bhabha National Institute, Kolkata, India}
		\address[label2]{CERN,CH-1211 Geneva 23, Switzerland}

\begin{abstract}

The scheme of the data acquisition (DAQ) architecture in High
Energy Physics (HEP) experiments consist of data transport from the
front-end electronics (FEE) of the online detectors to the readout units (RU), which perform online processing of the data, and then to the data storage for offline analysis.  With major upgrades of the Large Hadron Collider (LHC) experiments at
CERN, the data transmission rates in the DAQ systems are
expected to reach a few TB/sec within the next few years.
These high rates are normally associated with the
increase in the high-frequency losses, which lead to distortion in the
detected signal and degradation of signal integrity.
To address this, we have developed an optimization technique of the multi-gigabit
transceiver (MGT) and implemented it on the state-of-the-art
20nm Arria-10 FPGA manufactured by Intel Inc. 
The setup has been validated for three available high-speed data
transmission protocols, namely, 
GBT, TTC-PON and 10 Gbps Ethernet. The improvement in the signal
integrity is gauged by two metrics, the Bit Error Rate (BER)
and the Eye Diagram. It is observed that the technique improves the
signal integrity and reduces BER.  The test
results and the improvements in the metrics of signal integrity for
different link speeds are presented and discussed.

\end{abstract}

%\pacs{29.85.Ca}% PACS, the Physics and Astronomy
%% Classification Scheme.
%\keywords{DAQ, Transceiver, high-speed data communication, FPGA, Signal Integrity, Eye Diagram}%Use showkeys class option if keyword
%%display desired
%\maketitle
\end{frontmatter}

\section{Introduction} 

The major goals of HEP
experiments are to probe the fundamental constituents of the matter and understand the nature of 
fundamental forces. 
Advanced research in HEP demands a progressive increase in
collision energies and beam luminosities of the particle accelerators,
which are essential for accessing rare probes with extremely low 
cross sections~\cite{morrissey2012physics}.
The experiments are continuously upgraded with sophisticated
detectors, electronics and DAQ systems~\cite{panofsky1997evolution,smith2004trigger}. 
The DAQ architectures have been evolving continuously 
to cope up with the demands of the experiments~\cite{khan2017potent,toledo2003past}.
The LHC at CERN will go through a major upgrade during the long
shutdown (LS2) period, following which
the beam luminosities will increase by about an order of magnitude
from their present values. 
At the same time, the experiments at the LHC are 
upgrading the detector and DAQ systems to allow for faster readout of the online data.

The DAQ architecture in HEP experiments consists of the three general steps: (i) the data from the online detectors are transferred 
to the FEE through the detector backplane, (ii) the data from the FEE
are transferred to the RU~\cite{mitra2016common,gutierrez2005alice,khan2018development}, 
and (iii) the processed data are further transferred to data storage. 
These steps require high-speed data communication links from one step to the other. Most of the DAQ systems are designed using the present available technology in such a way that it could be easily upgraded to match the requirements of the system.
Since one of the major concerns is to efficiently acquire data for 
all the collisions, 
error resilient and efficient data transmission with minimal signal attenuation is required. Signal integrity is essential for
the proper Clock and Data Recovery (CDR)~\cite{mitra2016common,razavi2002challenges}.
%To avoid frequent replacement of the DAQ infrastructure and for an economical upgrade of the present systems, experiments settle for supporting the legacy systems at the rates beyond their speculated values~\cite{mitra2016common,CRU_TDR}.
Thus it is a challenge to minimize the bit error ratio (BER) and improve signal integrity for increased data rates~\cite{hall2011advanced}. 

In this manuscript we address the challenges of high-frequency losses
arising due to 
the high data rates for the DAQ systems in HEP experiments. 
Using FPGA we present a heuristic optimization technique to tune the
parameters of multi-gigabit transceivers for achieving the best performance at high-speeds for the transmission of data, trigger, timing and slow control information. 
%It allows the experiments to operate the legacy systems with stepwise upgrades. 
The proposed technique helps to improve the system performance in terms of signal integrity and is implemented
%The technique is implemented 
on a state-of-the-art 20nm Intel Arria-10 FPGA~\cite{MGTA10}. 
It uses the Intel-Altera on-die Instrumentation tools~\cite{TTK} and
does not require the probing of FPGA pins or transceiver attributes. The full setup is tested
for the link rate of the high-speed communication protocols frequently
used for data transmission in these experiments. 
The technique is useful for on-field system-level debugging, and the
parameters can be reconfigured dynamically, 
allowing the user to configure the transceivers for optimum
performance. The robustness of the optimization technique has been
tested with Pseudo Random Binary Sequence31 (PRBS31) pattern, which represents the stressed and transitional data conditions. 
For the statistical reliability of the performed tests, a large number
of data vectors are acquired.
Different performance indicators, such as, BER and eye diagrams
have been used to verify the improvement of
the quality of data signal posterior to the execution of proposed
optimization technique. 

The manuscript is organized as follows. In section~\ref{secAPU}, we present the data aggregation and processing in HEP experiments. 
The important constituents of the high-speed DAQ system are discussed in section~\ref{HSprotocol}. 
Details of the transceiver optimization technique with its intricate
features are presented in section~\ref{transceiveropt}. 
Section~\ref{testsetup} describes the FPGA based test setup, 
and section~\ref{method} discusses the methodology to implement the proposed technique and its advantages. 
The test results are presented and discussed in section~\ref{sec:results}. 
The manuscript is summarised in section~\ref{summary}.

\section{Data aggregation and processing}
\label{secAPU} 

\begin{figure}[!th]
	\centering 
    \includegraphics[width=\linewidth]{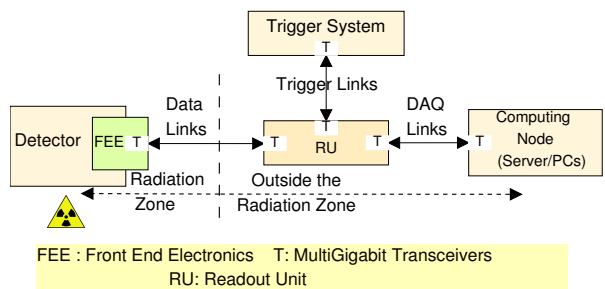}
	\caption{Basic blocks of a typical data acquisition architecture for HEP experiments.}
	\label{fig:Daqscheme}
\end{figure}

A generalised architecture for the DAQ scheme of the HEP experiments is presented in Figure~\ref{fig:Daqscheme}. 
The FEE boards are connected to the detectors and are located in the radiation zone with proximity to the 
detector, requiring custom-built radiation hard electronics. The FEE boards process the analog detector 
signals and convert those to digital signals. Design and specifications of these boards are unique to the 
individual detector system~\cite{knoll2010radiation}. The
particle detectors operate in the harsh
radiation zones and in some cases, in high magnetic fields. The main
data storage units, on the other hand, are kept in low radiation zones. The RUs, which
are intermediary between FEE and storage, can be placed either in the
radiation zone of the experiment's cavern or in a low radiation zone
near the data storage units. In an ideal case, the placing of the RUs near the
detectors in the cavern minimizes the transmission latencies. But it requires
custom-built radiation hard electronics, which are difficult to obtain.
In order to minimize the
effect of radiation, the RUs as well as the trigger system and the
back-end computing nodes, are kept out of the radiation zone. This
helps to get the advantage of the high processing power available
electronics with a large ecosystem, ease of
accessibility and maintenance.

The RU acts as an interface between detector data links, the trigger system, and links to 
storage as well as computing nodes as shown in Figure~\ref{fig:Daqscheme}. 
The tasks performed by the FPGA based RUs depend on the detector 
specifications and requirements. Main tasks are data sorting, optical link handling, multiplexing 
and forwarding of data from different interfacing links, embedding control and trigger information, 
etc.~\cite{li2018parallel}.
These versatile functionalities require RU to be designed on custom electronics boards 
with re-programmable functionality~\cite{castillo2005field}. It is based on up-to-date FPGA technology 
with embedded on-chip transceivers. For our tests we have used the Intel Arria-10 GX FPGA based development board~\cite{MGTA10, A10userguide}. The interfacing links of RU and the high-speed 
communication protocols used for the LHC experiments in the context of
the present framework 
are discussed in the following sections. 

\section{High-speed protocols}
\label{HSprotocol}
The DAQ architecture in Fig.~\ref{fig:Daqscheme} features three
different interfacing links: (i) the Data link, which connects the
detector FEE to RU, 
(ii) the Trigger link, which connects the RU to the 
trigger system of the experiment, and (iii) the DAQ link, which takes the data from the RU to the storage and 
computing nodes. 
For the data link, the Gigabit Transceiver (GBT) protocol
architecture~\cite{marin2015gbt}, developed at CERN, has been found to
be most ideal. The GBT protocol supports 
4.8 Gb/sec data transmission rate. It ensures the transmission of data from the FEE near the detectors in high radiation zone to the RU, which is located near the counting room in a low or no radiation zone.
%ensures fixed, low and deterministic latency.
%An improved protocol for the Data link has recently been proposed, which supports a frequency of 8.192 Gbps~\cite{mitra2016error}.
The Trigger link uses the
Timing, Trigger and Control system based on Passive Optical Networks (TTC-PON) technology~\cite{mendes201710g}; operates at the rate of 9.6 Gigabit per second. It ensures fixed, deterministic latency and satisfies the timing specification of the LHC. 

The data packets get time-stamped in the RU. Thus the links from the
RU to the computing nodes is not latency critical. It has been found
that the latest promising technology option of 10-Gigabit Ethernet~\cite{toledo2003past} with ample ecosystem are most suitable for the DAQ links in the experiments. In Table~\ref{table:HS_protocol}, we give the detailed specifications of the three interface links used in the HEP experiments for the acquisition of data. 
%It has been found
%that the 10-Gigabit ethernet and PCIe express 
%protocols~\cite{pci2002express} with the current releases 
%from the Peripheral Component Interconnect Special Interest Group (PCI-SIG) 
%are the latest promising technology options available with ample
%ecosystem, which are most suitable for the experiments~\cite{toledo2003past,bellato2014pcie,RSI2006ethernet}.
%In Table~\ref{table:HS_protocol}, we give the detailed specifications
%of the different interface links. 
\begin{table*}[ht!]
	\centering
	\caption{Specifications of three high speed interface links, GBT~\cite{marin2015gbt}, TTC-PON~\cite{mendes201710g} and 10-Gb Ethernet.}
	\label{table:HS_protocol}
	\scalebox{0.83} {
%	\resizebox{0.7\textwidth}{!}{%
		\begin{tabular}{@{}llll@{}}
\toprule
\textbf{Parameters}                                                             & \textbf{GBT}                                                                              & \textbf{TTC-PON}                                                                                    & \textbf{\begin{tabular}[c]{@{}l@{}}10Gb Ethernet\end{tabular}}           \\ \midrule
\textit{Technology Specification}                                               & Custom                                                                                     & \begin{tabular}[c]{@{}l@{}}XGPON1 with \\ modifications\end{tabular}                                & \begin{tabular}[c]{@{}l@{}}802.3ae Specification \\ Standard\end{tabular} \\ \midrule
\textit{Designer Group}                                                         & CERN                                                                                       & \begin{tabular}[c]{@{}l@{}}ITU-T with \\ CERN modifications\end{tabular}                            & IEEE                                                                        \\ \midrule
\textit{Line Rate}                                                              & 4.8 Gbps                                                                                   & \begin{tabular}[c]{@{}l@{}}Downstream: 9.6 Gbps\\ Upstream: 2.4 Gbps\end{tabular}                   & \begin{tabular}[c]{@{}l@{}}10.3125 Gbps per lane\end{tabular}            \\ \midrule
\textit{Payload Rate}                                                           & 3.2 Gbps                                                                                   & \begin{tabular}[c]{@{}l@{}}Downstream: 7.68 Gbps\\ Upstream: 640 Mbps\end{tabular}                  & \begin{tabular}[c]{@{}l@{}}10 Gbps per lane\end{tabular}                 \\ \midrule
\textit{Payload Size}                                                           & 120 bits@40 MHz                                                                            & \begin{tabular}[c]{@{}l@{}}Downstream: \\ 192 bits@40 MHz\\ Upstream:\\ 16 bits@40 MHz\end{tabular} & 64 bits@156.25 MHz                                                          \\ \midrule
\textit{Wavelength (nm)}                                                        & \begin{tabular}[c]{@{}l@{}}850 nm\\  (Multi-mode)\\  1310 nm \\ (Single-mode)\end{tabular} & \begin{tabular}[c]{@{}l@{}}Downstream: 1577 nm\\ Upstream: 1270 nm\end{tabular}                     & \begin{tabular}[c]{@{}l@{}}850 nm(10 Gb BASE-SR)\end{tabular}            \\ \midrule
\textit{Network Topology}                                                       & Point-to-Point                                                                             & Point-to-Multipoint                                                                                 & Point-to-Point                                                              \\ \midrule
\textit{Encoding}                                                               & \begin{tabular}[c]{@{}l@{}}RS ECC with Block \\ Interleaver\end{tabular}                   & 8b/10b                                                                                              & 64b/66b                                                                     \\ \midrule
\textit{\begin{tabular}[c]{@{}l@{}}Synchronous Trigger \\ Support\end{tabular}} & Yes                                                                                        & Yes                                                                                                 & No                                                                          \\ \midrule
\textit{Trigger Latency}                                                        & \begin{tabular}[c]{@{}l@{}}150 ns\\ (Optical loop-back)\end{tabular}                       & \begin{tabular}[c]{@{}l@{}}100 ns Downstream \\ 1.6 us Upstream\end{tabular}                        & NA                                                                          \\ \bottomrule
\end{tabular}
}
\end{table*}
\section{Transceiver optimization}
\label{transceiveropt}

High-speed data communication suffers from the transmission
losses and signal integrity issues; not seen at normal digital
signalling levels~\cite{hall2011advanced}. The
high-frequency content of the signal gets degraded due to dielectric
losses, skin effect, discontinuities in connectors, reflections caused
by the vias, inadequately placed traces, etc. 
We have developed a technique to optimize the transceiver 
parameters accurately and offer the best combination for a 
given high-speed link. This optimization of the transceiver
parameters could take care of the transmission losses~\cite{MGTSV}.

For the high-speed transmission channels with multi-gigabit 
rates, the unit interval (UI) for the data bit decreases.  
At high transmission rates, the PCB materials suffer 
from frequency dependent losses, hence become dispersive. 
This prevents the 
signal from reaching its full strength at the shrunk UI window,
leading to jitter and intersymbol interference (ISI). It also disturbs 
the deciphering of the signal and the extraction of the embedded clock becomes 
difficult at the receiver end.

An increase of the signal strength is an obvious solution to overcome the 
attenuation. However, the issue of high-frequency roll-off remains,
and the pattern dependent jitter gets aggravated. Consequently, the signal does not 
reach its optimal strength within the interval and may 
diffuse further into the next UI leading to ISI. Also for the increase
of signal strength overall power consumption of the transceiver
increases. Noise 
levels in the system also increase proportionally.
All these lead to deteriorated metrics of signal integrity and 
reduced drive length. The effects are even more evident with the use of
high-speed interfaces with the systems which were originally designed for low bandwidth applications. 

To overcome these losses, we have developed the
transceiver optimization technique and a
proficient methodology for 20nm Arria-10 FPGA. This new FPGA with
considerably large
on-chip resources~\cite{MGTA10} are ideal for the processing requiremnts in the experiments.

\subsection{Optimization Technique}
\label{sec:opt_tech}

For the optimization, the high-frequency components in the data
stream are boosted up on every switching, using the digital
pre-emphasis taps of the on-chip transceiver. In addition, the low
frequency components are reduced. This technique helps to achieve the same amount
of emphasis with less power dissipation. The exaggerations are
overridden by the attenuation during transmission and allow for
the signal to be recovered accurately. 

\begin{figure}[!th]
	\centering 
	\includegraphics[width=\linewidth]{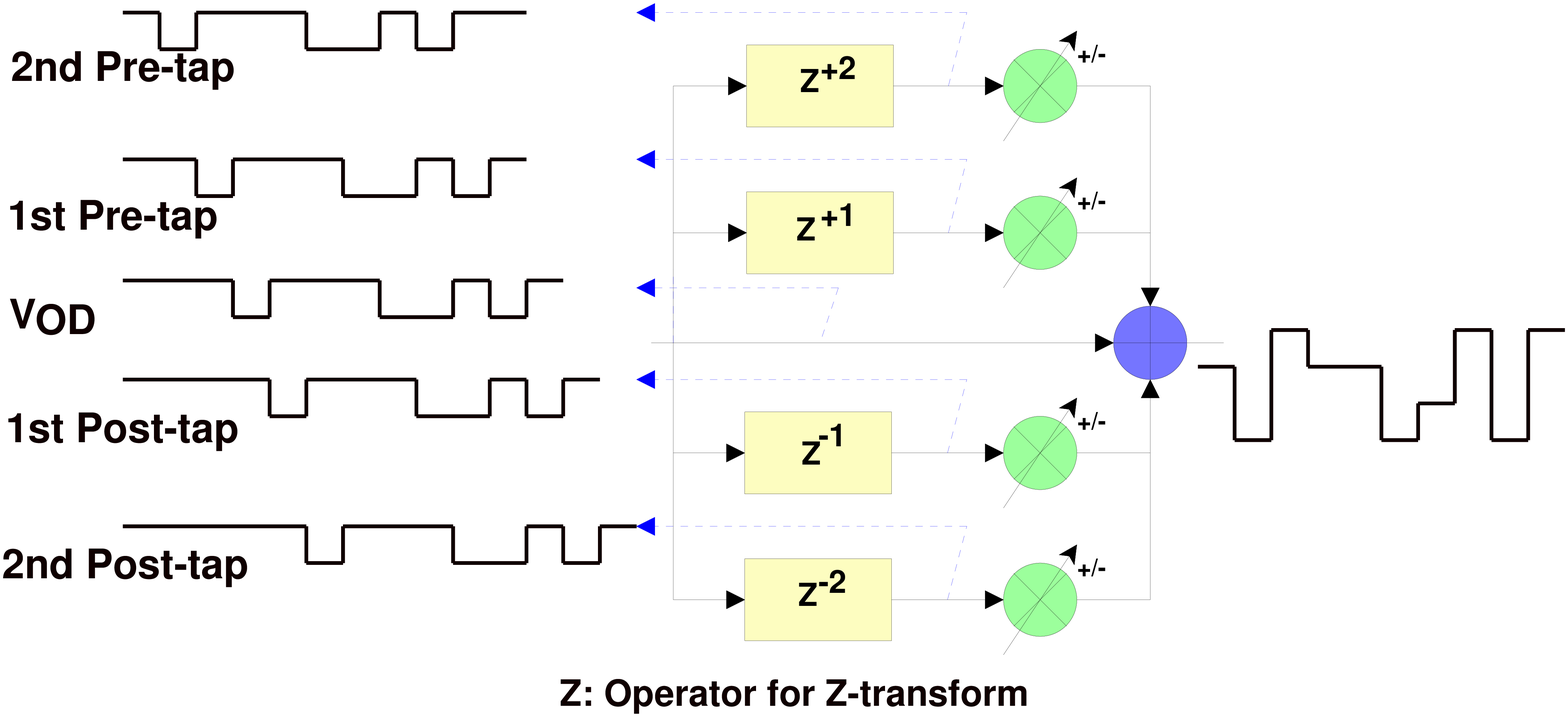}
	\caption{Voltage output differential (VOD) and tunable pre-emphasis taps with flexible polarity in the embedded transceiver of FPGA.}
	\label{fig:Preemp}
\end{figure}

The optimization technique has been implemented on Intel Arria-10
FPGA development board with integrated
reconfigurable transceiver
architecture~\cite{MGTA10}. It incorporates additional
circuitry in buffers for equalisation and pre-emphasis techniques. The
transmitter of the embedded transceiver has five programmable drivers
as shown in Figure~\ref{fig:Preemp}. Voltage output differential
(V{\small OD}) controls the base amplitude. The four pre-emphasis taps
are 1st pre-tap, 2nd pre-tap, 1st post-tap and 2nd post-tap. These taps also include polarity settings. The post taps are the causal taps and the pre-taps are the anti-causal taps. These multiple taps and choice of polarity could
handle channel attenuating
characteristics. Equalisation with DC gain and Variable Gain Amplifier
(VGA) is on the receiver side of the transceiver.
There are multiple transceiver parameters with a large span of operating range
and so to scan the system performance for every combination of the
parameters is a time-consuming process.
Our goal had been to develop an efficient technique for
optimization of transceiver parameters
such that the signals impacted by the high-frequency losses are recovered.

\begin{figure}[ht!]
	\centering
	\includegraphics[trim=0.0cm 0.0cm 0.7cm 0.0cm, clip=true, width=1.0\linewidth]{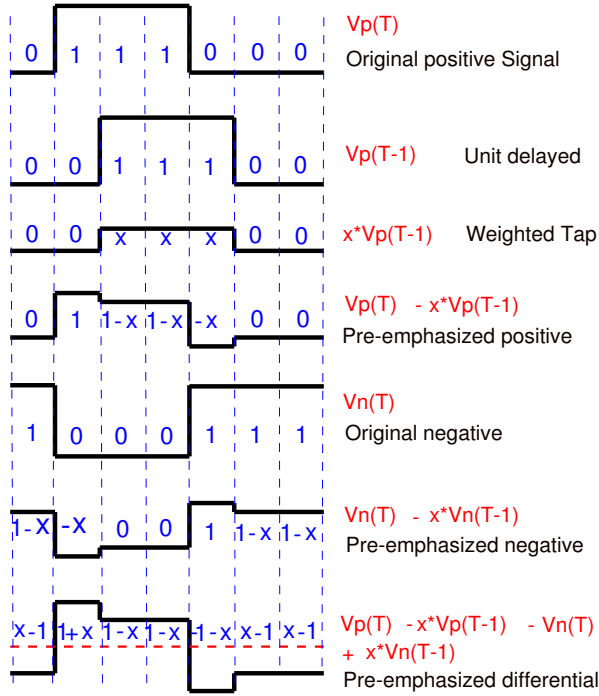}
	\caption{The pre-emphasis signal generation technique at the 1st post-tap in embedded FPGA transceivers, $0 <{\large x}<1 $ is the tap weight.}
	\label{fig:posttap1}
\end{figure}

It works like a \textit{Finite Impulse Response (FIR)} filter with different delays referred to as the taps as
shown in the Figure~\ref{fig:Preemp}. An FIR filter is based on a feed-forward difference equation. The pre-emphasis technique 
applies a delay to the signal and adds it back to the real signal with weight and inversion as and when required.  
Although depending on the transmission channel peculiarity, a simple
delay, weight and inversion 
may not be able to provide the required compensation. For this reason,
a combination of different delays, weights 
and the polarity are combined.    % {\bf explain Fig. 2 in few more lines: Answer; it is already explained in the two paragraphs before } 
In this configuration, the pre-emphasis 1st post-tap is the most useful parameter. It
emphasises the immediate bit period after the transition. The
generation of the differential emphasised signal, applying the unit
delay by the first post-tap is shown in Figure~\ref{fig:posttap1},
assuming V{\small OD}~=~1 and tap weight as $0 <{\large x}<1 $. The original positive signal Vp(T) is compared with Vp(T-1) which is the unit-delayed signal. The
emphasised signal is the difference between the weighted x*Vp(T-1) signal and the Vp(T) signal. The negative signal is similarly generated. The pre-emphasised differential signal is differentiated from the positive and negative signals.% {\bf describe Fig. 3}
\begin{figure}[!th]
	\centering
	\includegraphics[width=1.032\linewidth]{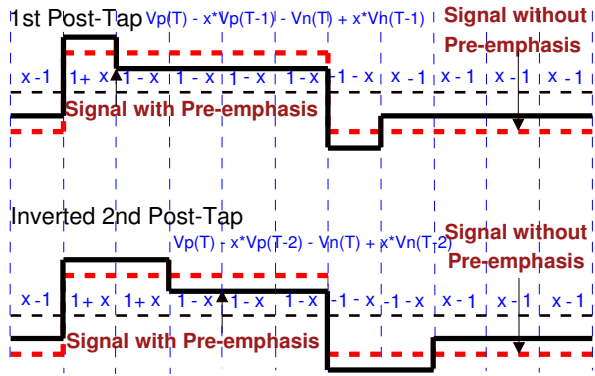}
	\caption{Pre-emphasis 2nd post-tap (Inverted) compared with pre-emphasis 1st post-tap and their effect on the signal without pre-emphasis.}
	\label{fig:posttap2}
\end{figure}
\begin{figure}[ht!]
	\centering
	\includegraphics[width=1.0\linewidth]{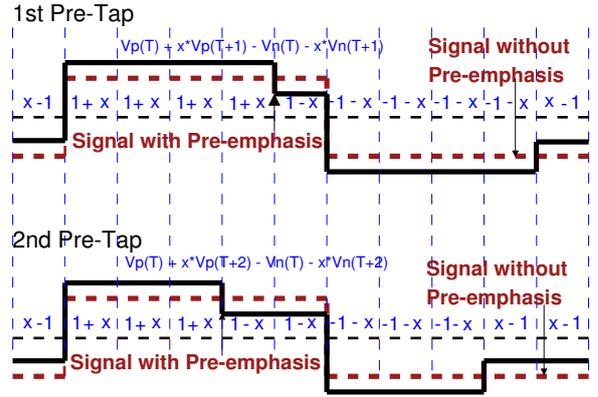}
	\caption{Pre-emphasis 1st pre-tap and the 2nd pre-tap (Inverted) and their effect on the signal without pre-emphasis.}
	\label{fig:pretap}
\end{figure}The effect of 2nd post-tap after the transition, 
depending on the chosen polarity setting is shown in Figure~\ref{fig:posttap2}.

The pre-tap reduces the effect of pre-cursor ISI. 
Figure~\ref{fig:pretap} shows 
the impact of 1st 
 pre-tap and the 2nd pre-tap on the single and double bit period
 respectively, before the occurrence of high-frequency transition depending on the 
 polarity. Both pre-cursor ISI and post-cursor ISI are handled by
anti-causal and causal taps respectively. However, pre-emphasis alone
cannot guarantee the performance of the system as it is implemented at
the transmitter by pre-conditioning the signal before it is fed
to the channel. There are high-frequency losses in the transmission
channel itself. Hence an equalisation is required at the receiver
end. It compensates for the low pass characteristics of the physical medium
and amplifies the attenuated high-frequency components of the incoming
signal. An equalizer on the receiver side lifts the contents inside a band of
frequencies and attenuates the rest. The DC gain circuitry gives uniform
amplification to the received spectrum. It enables the transceivers to
operate over longer distances. The VGA on the receiver optimizes the
signal amplitude before the CDR sampling.  
%{\it MAKE SIMPLE: It seems many things are repeated from before:
%The real transmission link does not represent a simple attenuation curve. There are reflections, crosstalks, resonances at some frequencies and multiple poles in the transfer function. It demands distinct settings for drive strengths, equalisation and pre-emphasis. 

To achieve an optimal signal integrity performance, both transmitter
and receiver parameters of the transceiver on 
FPGA chip augments each other and work combined to compensate for the high-frequency losses.
However, the overcompensation degrades the signal quality and adds more jitter leading to the closed eye diagram rendering it futile for the receiver to identify the signal and hence should be avoided.

\section{Test setup}
\label{testsetup}

 \begin{figure*}[ht!]
 	\centering 
 	\includegraphics[trim=0.0cm 0.0cm 0.0cm 0.0cm, clip=true, width=0.85\linewidth]{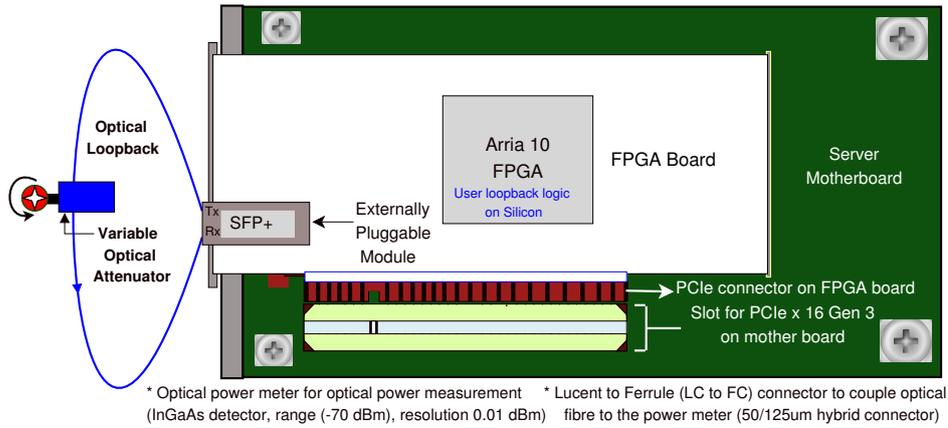}
 	\caption{%FPGA based test setup with Arria-10 FPGA card inserted in PCIe x16 slot of server. 
%for BER tests and optimization of embedded transceiver. 
       Arria-10 FPGA card inserted in PCIe x16 slot of server. 
The optical signal from the externally pluggable SFP+ is looped back 
via the fibre equipped with the variable optical attenuator (VOA). }
 	\label{fig:BER_test}
 \end{figure*}

An FPGA based setup has been developed to test the potency of the proposed optimization technique. The transceiver is tested for the high-speed links under the stressed conditions. The setup has been utilised to emulate the stressed high-speed link conditions and to investigate the high frequency losses in the transmission. It determines the capability of the transceiver system to recover the data from the degraded signals. Tests are performed at the system level to operate the setup at a prescribed BER equal to or better than $10^{-12}$ as per the IEEE standard.

The test setup, shown in Fig.~\ref{fig:BER_test}, engrosses the
 Arria-10 FPGA development board (10AX115S2F45I1SG device) for the
 implementation and testing of the optimization technique. 
The FPGA development card is installed on the PCIe 16 lane slot of the server, where the power is obtained from the  server motherboard. 
The functions and specifications of each of the components of the setup are given in Table~\ref{table:system}.

	\begin{table*}[!t]
		\centering
		\caption{Components used in the test setup, their role and specifications.}
		\label{table:system}
		\scalebox{0.92} {
\begin{tabular}{lll}
	\hline
	\textbf{Component}                                                                              & \textbf{Role in test setup}                                                                                                                                                                                                                                          & \textbf{Specification}                                                                                                                                                                                                                                                        \\ \hline
	FPGA Test Board                                                                                 & \begin{tabular}[c]{@{}l@{}}Integrated FPGA based design environment\\ with embedded transceivers on silicon. PCIe \\connection. Slot for hot pluggable transceiver \\optical modules. Other accessories \end{tabular} & \begin{tabular}[c]{@{}l@{}}Intel Arria10 FPGA, (20nm mid-range).\\ Transceivers upto 17.4 Gbps~\cite{MGTA10}. \end{tabular}                                                                                                         \\ \hline
	\begin{tabular}[c]{@{}l@{}}Variable Optical\\ Attenuator (VOA) with \\optical Fiber\end{tabular} & \begin{tabular}[c]{@{}l@{}}Optical power attenuation\\ in the fibre loopback path.\end{tabular}                                                                                                                                          & \begin{tabular}[c]{@{}l@{}}Range(dB)-0$\sim$60, Accuracy +/- 0.8dB.\\ Fibre(850nm): Multimode 50/125um with\\ Lucent connector (LC), Dia-2mm,\\ Insertion loss \textless{}2.5dB, Length-2 m\end{tabular} \\ \hline
	\begin{tabular}[c]{@{}l@{}}Serial Form factor \\ Pluggable (SFP+)\\ module.\end{tabular}        & \begin{tabular}[c]{@{}l@{}}External transceiver modules to be\\coupled to the fibre. Laser at transmitter \\and PIN diodes at the receiver ends \end{tabular}                                                 & \begin{tabular}[c]{@{}l@{}}Hot-pluggable footprint, upto 10Gbps, \\ 850nm VCSEL laser, duplex LC connector. \\ Link length of 300m~\cite{sffsff}. \end{tabular}                                                                        \\ \hline
%	Optical power meter                                                                             & \begin{tabular}[c]{@{}l@{}}To measure the received optical\\power output\end{tabular}                                                                                                                                                        & \begin{tabular}[c]{@{}l@{}}InGaAs detector, range (-70$\sim$+6dBm)\\ uncertainty 5\%, Resolution 0.01 dB.\\ $\lambda =  850nm$.\end{tabular}                                                                                      \\ \hline
%	\begin{tabular}[c]{@{}l@{}}LC to\\Ferrule connector(FC)\end{tabular}                      & \begin{tabular}[c]{@{}l@{}}To connect the fibre cable with \\ LC connector to the power meter.\end{tabular}                                                                                                                                                          & \begin{tabular}[c]{@{}l@{}}Multimode 50/125$ \mu $m hybrid connector.\\ Insertion loss \textless{}0.3dB at 850 nm\end{tabular}                                                                                                                                                      \\ \hline
	\begin{tabular}[c]{@{}l@{}}Workstation with FPGA \\ design platforms\end{tabular}               & \begin{tabular}[c]{@{}l@{}}FPGA board powered through PCIe \\Gen3x16 slot. Compile and \\generate the FPGA design with \\ firmware development softwares\end{tabular}                                                                               & \begin{tabular}[c]{@{}l@{}}PCIe Gen3 x16 slots available. Quartus-II \\platform installed for firmware design \\and generation. FPGA programmed through\\ USB blaster download cable. \end{tabular}                                                        \\ 
	%\hline
%	JTAG Cable                                                                                      & To load the generated design bit file to FPGA                                                                                                                                                                                                                        &USB type.\\ 
	\bottomrule
\end{tabular}}
\end{table*}

\begin{figure}[!th]
	\centering  
	\includegraphics[trim=0.1cm 0.1cm 0.1cm 0.1cm, clip=true, width=1.0\linewidth]{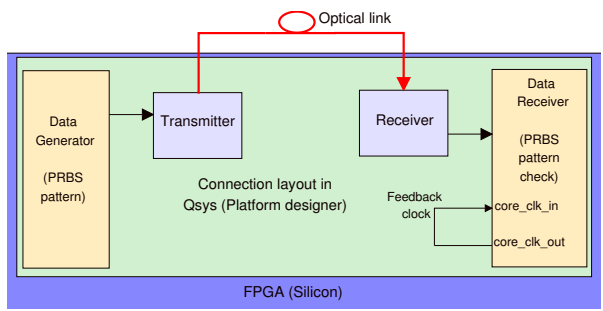}
	\caption{Typical BER test loopback logic on FPGA using Qsys tool.  The serialised data is transmitted, looped back and checked for the flipped bits at the receiver.}
	\label{fig:user_logic}
\end{figure}

Intel Quartus-II platform is the firmware application package,
implemented on the FPGA logic design. The transmission links at the 
specified data rates are implemented using Quartus-II Qsys tool.
Qsys is Intel${'}$s system integration tool for the quick generation
of the interconnect logic. The signal integrity of the transceiver
links is validated using Transceiver Toolkit (TTK) feature of
Quartus-II with a GUI. The TTK is used to quickly access, tune and test
the transceiver parameter settings in runtime through a combination of
metrics. The TTK enables us to measure BER and the eye diagrams and
also verify the signal integrity in external loopback mode. Details of
the firmware-tools, such as, Quartus II, Qsys, TTK, PRBS patterns and
auto-sweep features may be found in reference~\cite{TTK}.

For the data loopback tests~\cite{green2002multichannel}, multimode optical fibre equipped with 
Variable Optical Attenuator (VOA) and external pluggable SFP+ modules are used. 
The far end of the transceiver is coiled back to the receiving
end. The received data is then verified by the data checker logic on
FPGA for any erroneous bits as shown in
Figure~\ref{fig:user_logic}. To test the signal 
integrity a variety of data patterns can be used.
However, in each case, a checker must be available for verification.
PRBS patterns are injected into the test
system as it generates the stressed and lengthy patterns with fewer
memory consumption~\cite{badaoui2010prqs}. 
Another advantage of using PRBS patterns for the tests is that the
boundary synchronisation is not necessary at the physical layer as the 
patterns are time correlated. The Intel soft logic cores are used for PRBS data pattern generator and checker~\cite{TTK}.

The BER measurement approach was chosen with respect to the controlled
attenuated optical power at the receiver with the help of VOA. It
allowed us to rapidly characterise the transceiver sensitivity below
which the embedded clock cannot be recovered from the data stream, and
loss of lock occurs~\cite{MGTSV}. It also determines the minimum
required optical power to achieve the targeted BER for a system
operating at a specified data rate. Auto sweep feature of TTK is used
to obtain the optimum settings of the best performing parameters of
the transceiver for a specified BER. This optimized set of transceiver
parameters delivers the best metrics of
signal integrity and the eye diagrams by its height and width. 
In the next section, we elaborate the methodology for the optimization
of high data rate on-chip transceivers to reduce the effect of high-frequency losses.

\section{Methodology}
\label{method}

\begin{figure*}[ht!]
	\centering
	\includegraphics[width=0.7\linewidth, height=20.5cm]{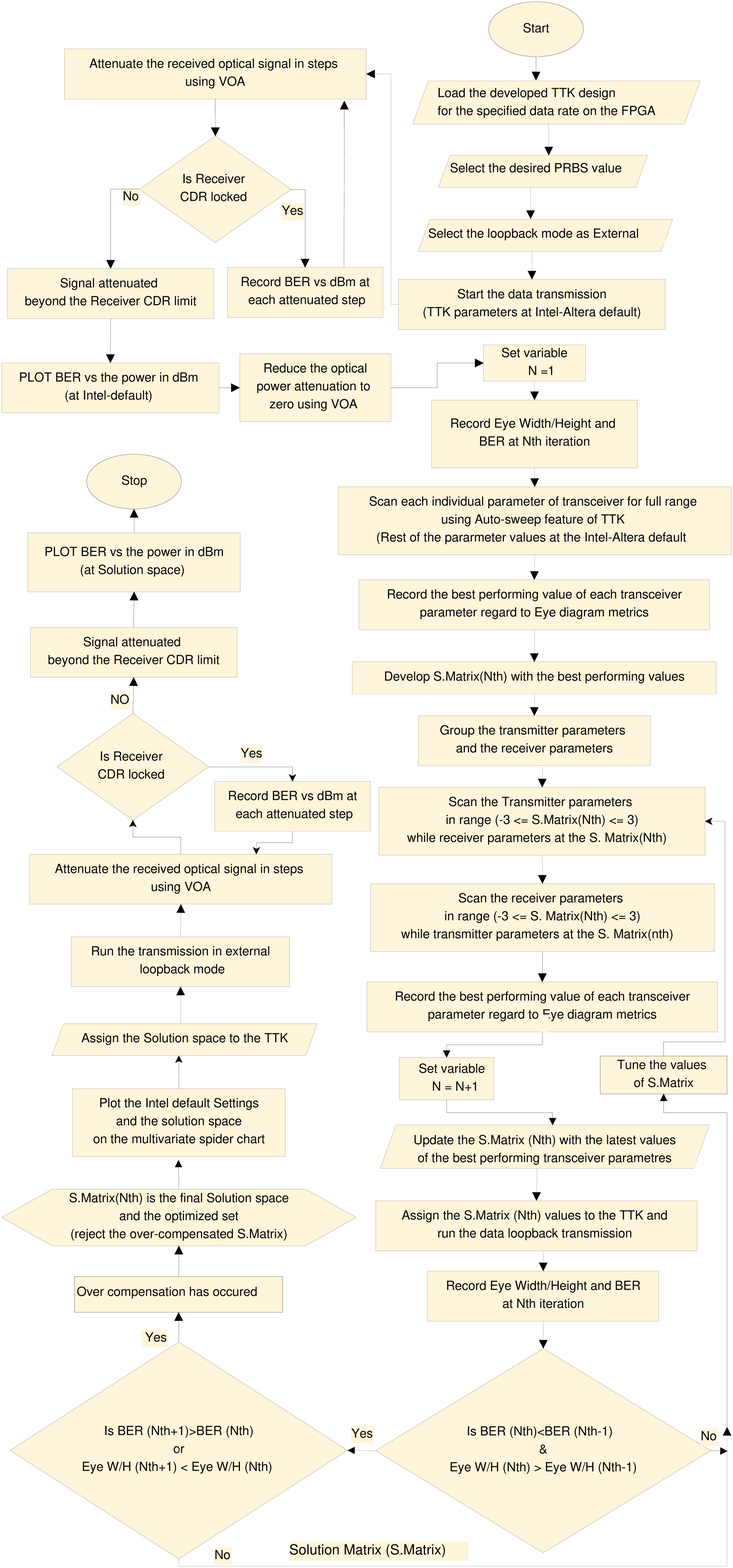}
	\caption{Stepwise flow diagram for the Transceiver Optimization. Data transmission is started with the Intel default parameters and a Solution matrix is derived to achieve the optimized signal integrity}
	\label{fig:flowchart}
\end{figure*}

The methodology to extract the optimized settings of the  
transceiver parameters has been explained in the flowchart in Figure~\ref{fig:flowchart}. 
To start with, the optimization process scans the full range of
each transceiver parameter using the TTK auto-sweep feature while the
rest of the parameters are set at their Intel-default values. Then it records
the best performing tap setting values for each transceiver parameter
as indicated by eye parameters. At this instance, a Solution
Matrix $\mathbb{(\textbf{S})}$ at $ Nth $ iteration, set $ N=1 $ is developed.
Then, we separately group the transmitter parameters viz. {V{\small OD}, Pre-emphasis (1st pre-tap, 2nd pre-tap, 1st post-tap, 2nd post-tap)}
and the receiver parameters (DC gain, Equalisation control,
  VGA). Then we scan again the transmission and receive parameters separately in
the range of -3 $\leq$ $\mathbb{\textbf{S}}$ $\leq$ 3, while receive
and transmit parameters respectively are set at the values enlisted in
the $\mathbb{\textbf{S}}$. Record again the best performing cases and
update the $\mathbb{\textbf{S}}$ with newer values, increment N by 1.
Assign the latest matrix values to the TTK and run the loopback
test. If this
does not result in the improved metrics of signal integrity (Eye diagram and the BER) than the one obtained at the
Intel default set values; repeat the optimization loop with
the adjusted $\mathbb{\textbf{S}}$ values in the range defined until
the improvement in both eye diagram and BER is achieved. 

The parameters cannot be declared as optimized until a stage of
degradation in the signal integrity metrics from their peak values is
observed. The degradation of metrics denotes the over-compensation and
it marks the transition from the maxima of the transceiver
parameters. Assign and update the $\mathbb{\textbf{S}}$ with the best
performing case metric values rejecting the over-compensated value
set. The final $\mathbb{\textbf{S}}$ values with the best performing
metrics is known as \textit{Solution Space}~\cite{MGTSV}. The deduced
final values are fed to the transceiver for further analysis. The results are presented and
discussed in the next section.

The proposed technique has definite advantages over traditional method
where the transceiver optimization may be carried out in an extremely time-consuming way by evaluating the signal integrity through a large
number of permutations and combinations of the parameters. The
parameters and their possible ranges are listed in the Table~\ref{table:advantagetime}.
\begin{table}[!th]
	\centering 
	\caption{Transceiver parameters, range of operations for the manual optimization.}
	\scalebox{0.92}{
		\label{table:advantagetime}
		\begin{tabular}{@{}ccc@{}}
			\toprule
			\textbf{\begin{tabular}[c]{@{}c@{}}Transceiver \\ parameter\end{tabular}} & \textbf{\begin{tabular}[c]{@{}c@{}}Range of \\ possible \\ values\end{tabular}} & \textbf{\begin{tabular}[c]{@{}c@{}}Number of \\ iterations \\ required\end{tabular}} \\ \midrule
			\multicolumn{3}{c}{\textit{\textbf{Transmitter Side}}}                                                                                                                                                                                             \\ \hline
			\textit{VOD}                                                              & 0 to 31                                                                         & 32                                                                                   \\ \hline
			\textit{Pre-emphasis 1st post-tap}                                        & -31 to 31                                                                       & 63                                                                                   \\ 
			\textit{Pre-emphasis 1st pre-tap}                                         & -31 to 31                                                                       & 63                                                                                   \\
			\textit{Pre-emphasis 2nd post-tap}                                        & -15 to 15                                                                       & 31                                                                                   \\
			\textit{Pre-emphasis 2nd pre-tap}                                        & - 7 to 7                                                                        & 15                                                                                   \\ \hline
			\multicolumn{3}{c}{\textit{\textbf{Receiver Side}}}                                                                                                                                                                                                \\ \hline
			\textit{DC gain}                                                          & 0 to 4                                                                          & 5                                                                                    \\
			\textit{Equalisation}                                                     & 0 to 15                                                                         & 16                                                                                   \\
			\textit{VGA}                                                              & 0 to 7                                                                          & 8                                                                                    \\ 
			\bottomrule 
	\end{tabular}}
\end{table}

\section{Results and discussion} 
\label{sec:results}

Results are demonstrated and validated for the three different high
speed optical links: 10 Gbps links, 4.8 Gbps GBT protocol and 9.6 Gbps TTC-PON. 
The test system confronts the lock and hold capability of the CDR circuit,
perturbs all the conceivable instances of ISI and analyses the
receiver sensitivity for any probable drifts. Drifts at the receiver
are caused due to long imbalanced runs of the data transition 
pattern. 
%\subsection{The eye diagram}
The PRBS31, $2^{31}-1$ patterns integrate every alteration of 31
bits. It gives a random sequence of bits with high and low
transitional values as defined by the logic levels of FPGA. The
different combinations induce non-similar ISI configurations. It is
required to stress the transceivers, test any innate ISI in a
transmitter, and to assess the quality of transmission. PRBS patterns
depict a white spectrum in the frequency domain and are injected to
tests the robustness of the high-speed links. 
For the entire analysis, PRBS31 is used to stress the system. However,
the variation of eye diagram and BER characteristics are also studied
for PRBS7, PRBS9, PRBS15, PRBS23 in addition to PRBS31.

\subsection{Eye Diagram analysis}

\begin{figure}[ht!]
	\centering
	\begin{tabular}{c}
		\includegraphics[trim=6cm 2.5cm 4cm 1.84cm, clip=true, totalheight=0.20\textheight]{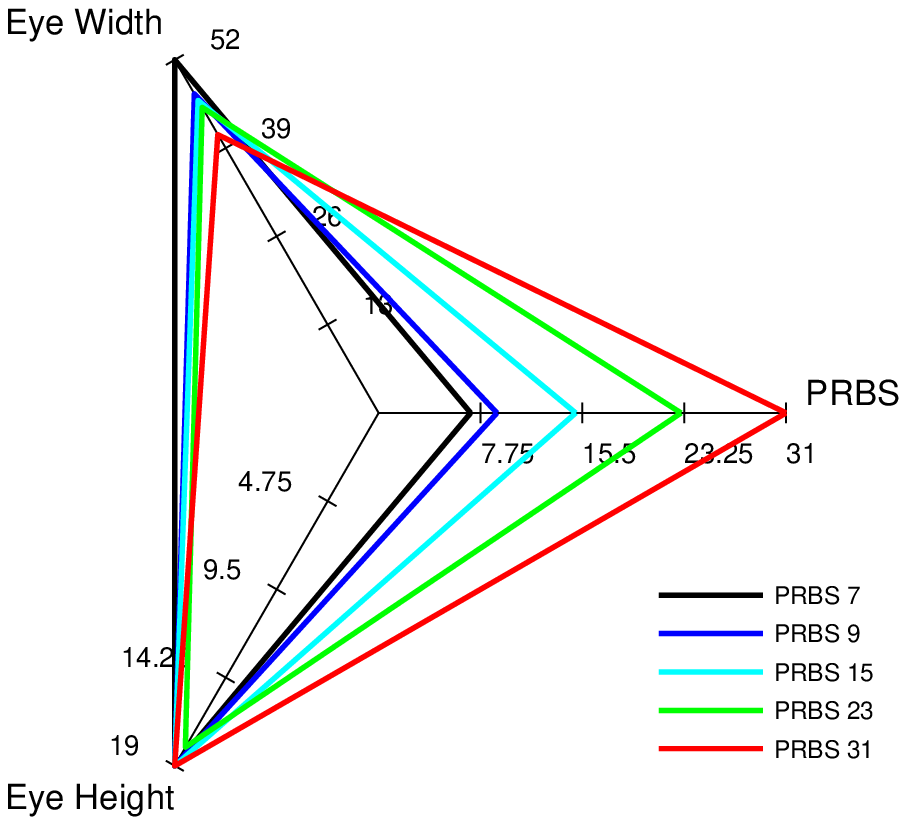}\\
		10 Gbps\\
		\includegraphics[trim=6cm 2.5cm 4cm 1.84cm, clip=true, totalheight=0.20\textheight]{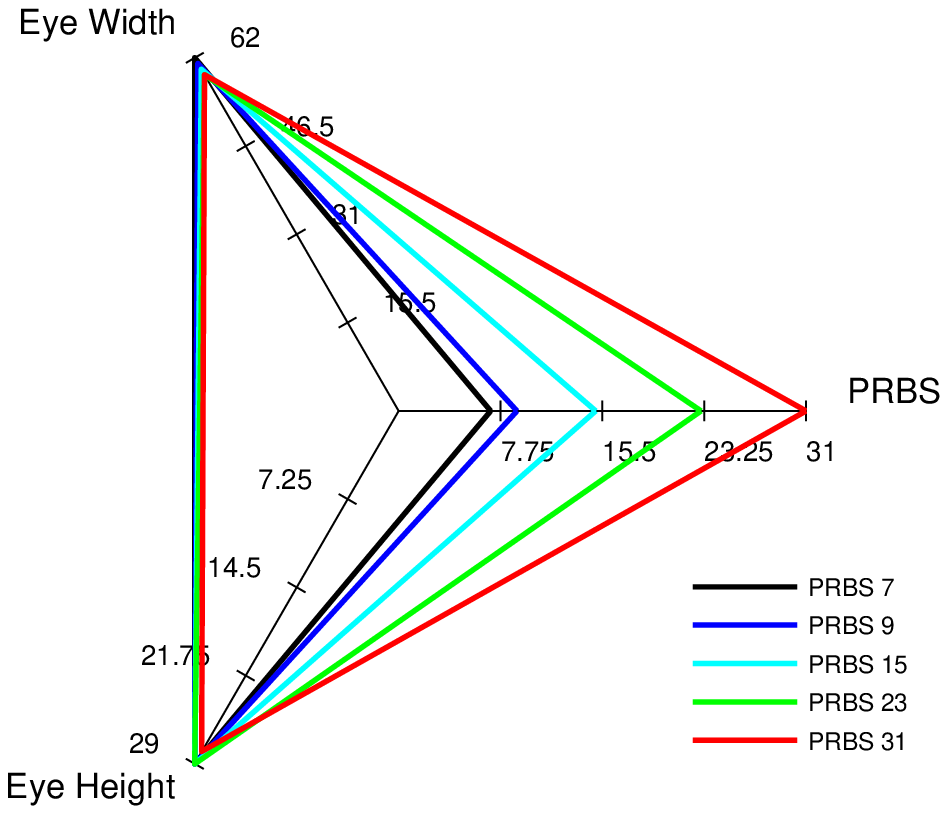}\\
		GBT 4.8 Gbps\\
		\includegraphics[trim=6cm 2.5cm 4cm 1.84cm, clip=true, totalheight=0.20\textheight]{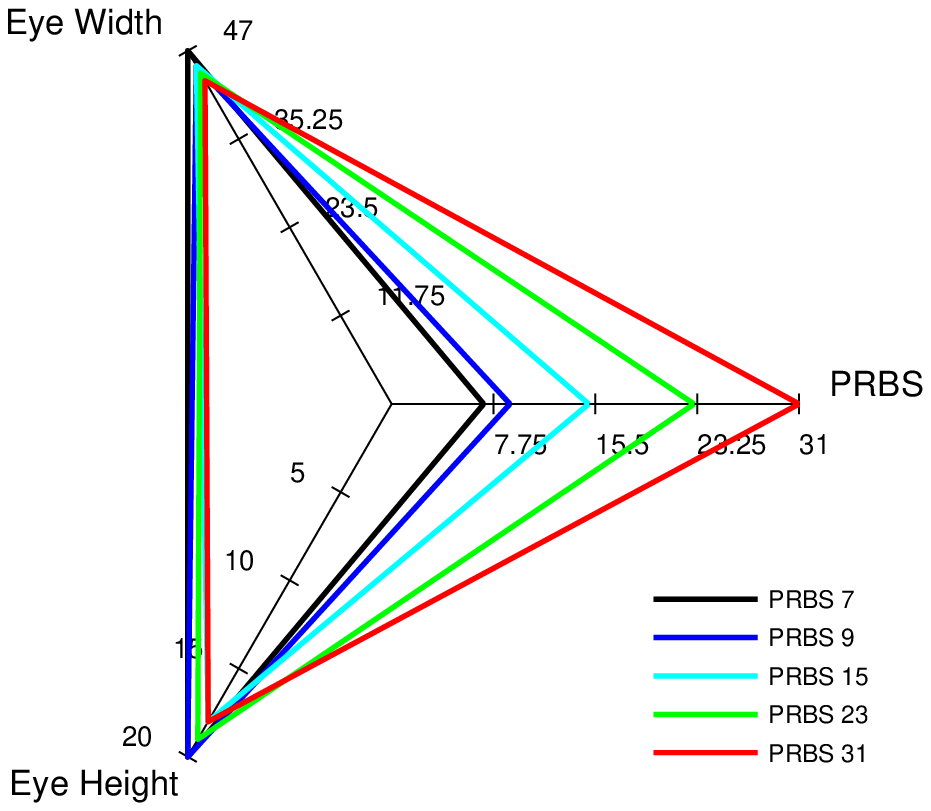}\\
		TTC-PON 9.6 Gbps
	\end{tabular}
	\caption{Changes in the Eye height and Eye width with PRBS variation for optical links at three line rates.}
	\label{fig:prbsall}
\end{figure}

At the system startup, the transceiver parameters in TTK are set at
the default values. Changes in eye diagram are compared for different
PRBS stressed patterns as the first set of analysis. Eye Height and
Width is plotted on a three axes plot with PRBS pattern on the third
axes as shown in Figure~\ref{fig:prbsall}. It is found that PRBS31 has
the most stressed eye metrics and as anticipated a more closed eye is
examined for all the three links speed. 

\subsection{BER Results}

Another important metric of signal integrity is BER. Its measurement
is a statistical phenomenon and the estimate is ideal only if the
number of tested bits tends to infinity, which is not possible in a
real lab test setup. Hence, a method was proposed in
reference~\cite{mitic2012calculating} to limit the stressing time
of a system to a feasible length and to measure the BER with high confidence level (CL) too. CL is used to quantify the quality of the estimate in
percentage. It is the system’s actual probability of error less than
the specified limit. The minimum number of bits required to be tested
for the BER measurement with a specific associated CL is given in
equation~\ref{eq:BERT}:
\begin{align}
\left. 
\begin{tabular}{rl}
$ n $ &$ = -\dfrac{\ln(1-CL)}{BER} + \dfrac{\ln\left(\sum_{k=0}^{N} \dfrac{(n*BER)^k}{k!}\right)}{BER} $\\
$ T $&$ =n/R $
\end{tabular}\right\}
\label{eq:BERT}
\end{align}
$ T $ is test time needed, $ R $ is the line rate and when $ N=0 $ the solution is trivial given in equation~\ref{trivial}. 
% when N = 0 
\begin{align}
\begin{tabular}{rl}
$ n $ &$ = -\dfrac{\ln(1-CL)}{BER} $
\end{tabular}
\label{trivial}
\end{align}
Where $ n $ are the total number of bits transmitted and $ N $ are the
number of errors that occurred during the transmission. There is a
compromise between testing time and the required accuracy of the
measurement as shown in equation~\ref{eq:BERT}. 

For the 95 percent CL, equation~\ref{trivial} reduces to n $\simeq 3/(BER)$. Hence to achieve the BER of $10^{-12}$ at 95 percent CL, total 3x$10^{12}$ bits need to be tested, as a thumb rule.

\subsubsection{BER analysis for various link speeds}

\begin{figure}[!th]
	\begin{center}
		\includegraphics[trim=0.9cm 0.55cm 1cm .66cm, clip=true, scale=0.295]{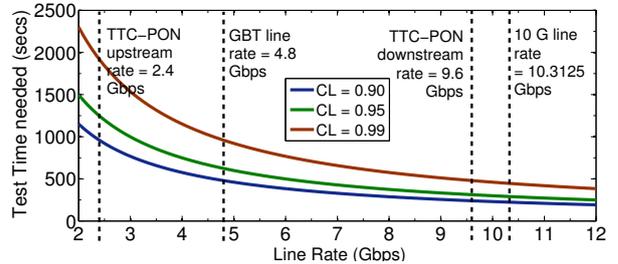}
		\caption{Time to achieve BER of $10^{-12}$ for the Line rate of GBT, TTC-PON and 10 Gbps optical links having different CL.}
		\label{fig:CL1}
	\end{center}
\end{figure}

\begin{figure}[t!]
	\centering
	\begin{tabular}{c}
		\includegraphics[trim=1.7cm 0cm 1cm 0cm, clip=true, scale=0.3]{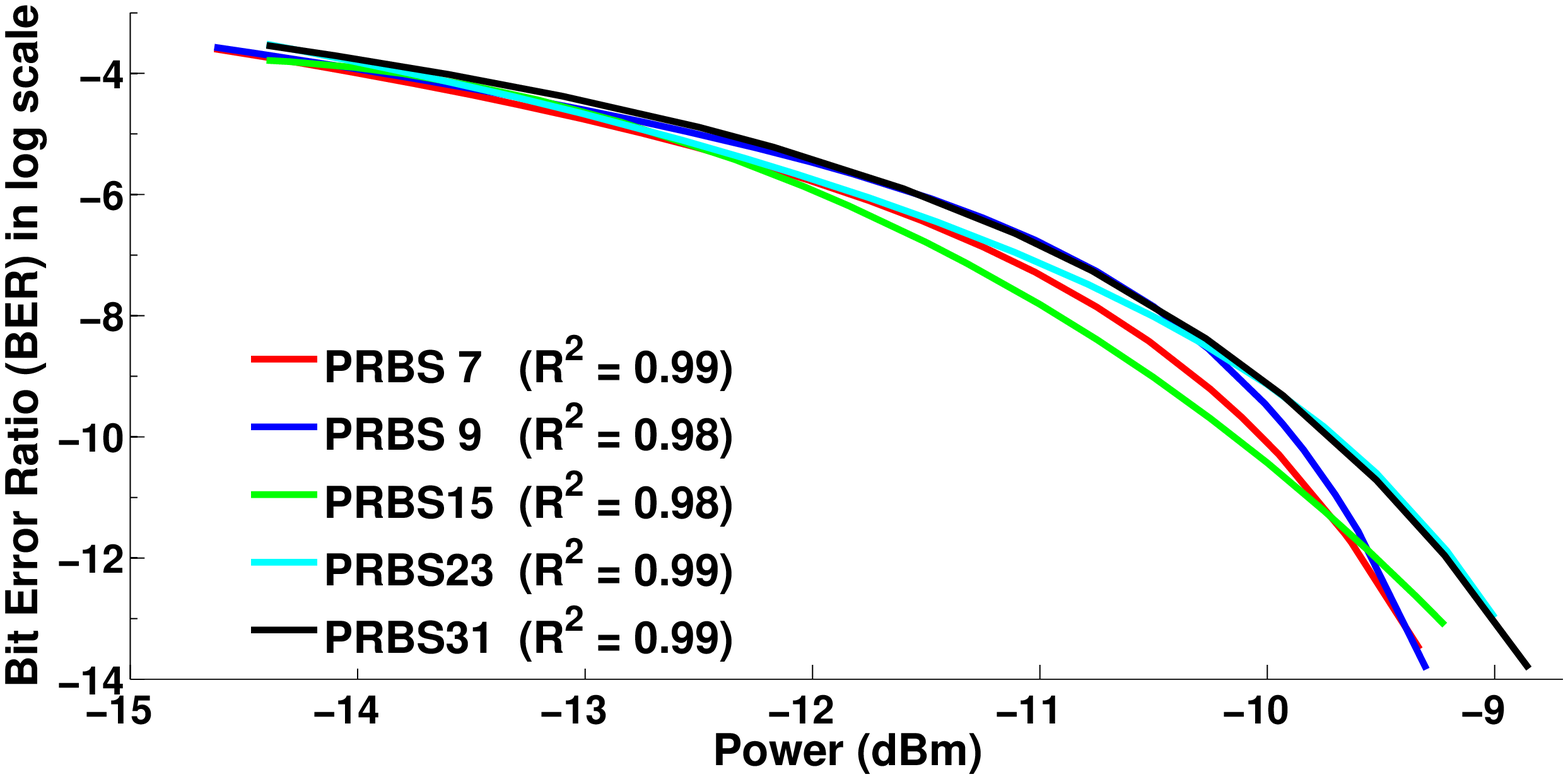}\\
		10 Gbps\\
		\includegraphics[trim=1.7cm 0cm 1cm 0cm, clip=true, scale=0.3]{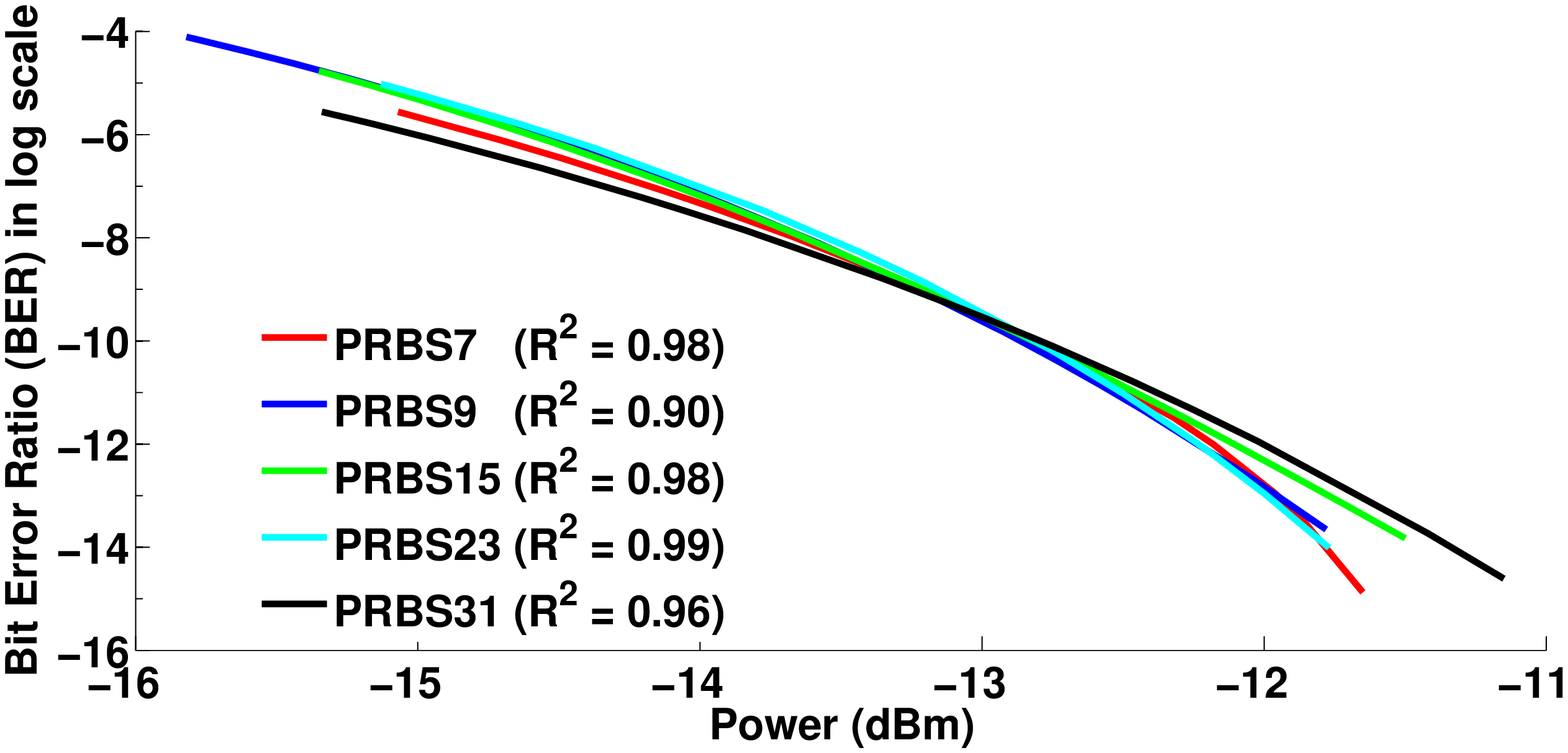}\\
		GBT 4.8 Gbps\\
		\includegraphics[trim=1.7cm 0cm 1cm 0cm, clip=true, scale=0.3]{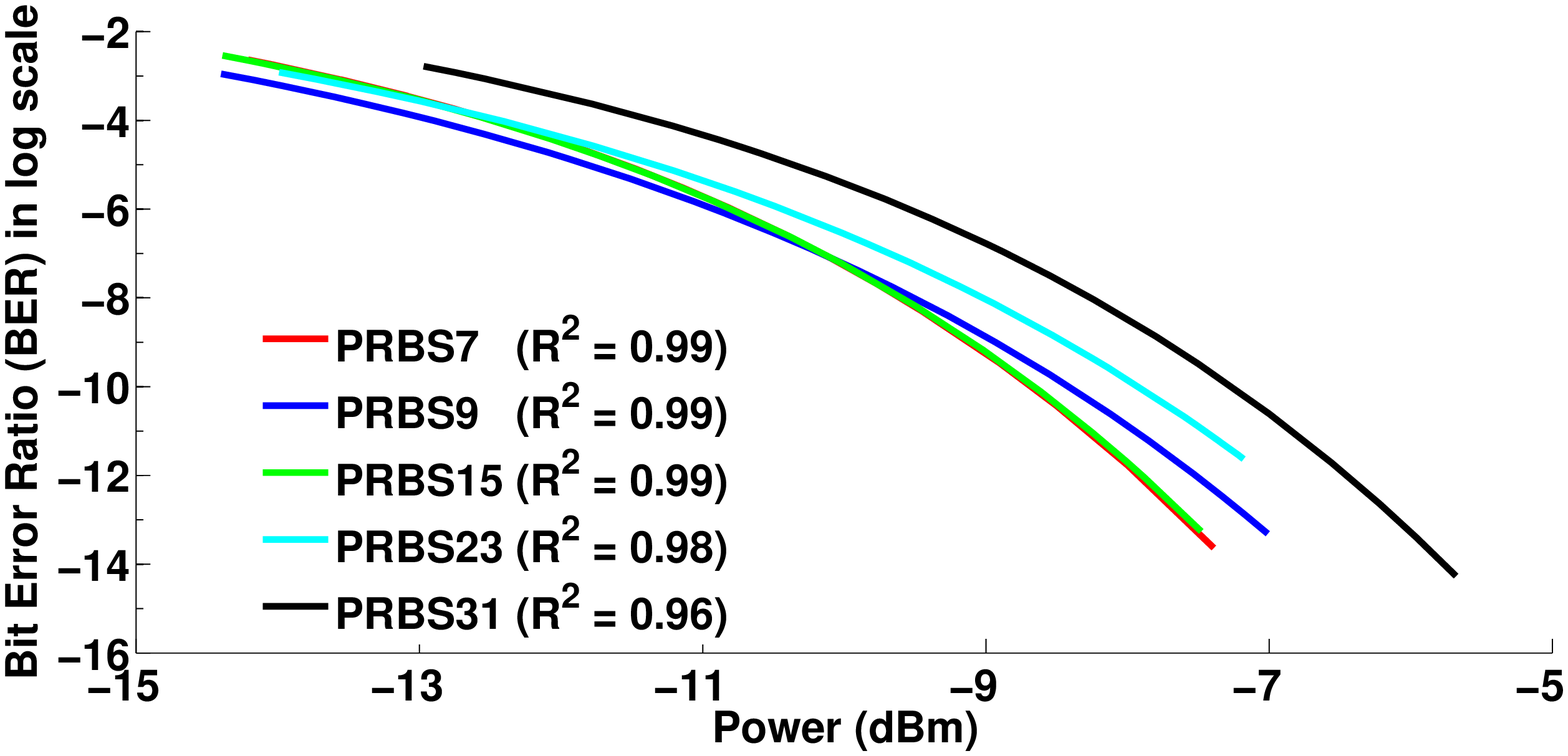}\\
		TTC-PON 9.6 Gbps
	\end{tabular}
\caption{BER versus received optical power(dBm) for transceiver at Intel FPGA default settings for different PRBS operating in three line rates.}
\label{fig:bervsdbm_all}
\end{figure}

The concept is further extended to find the minimum inspection time required to measure BER of $10^{-12}$ for different CL with no errors for GBT, TTC-PON and 10 Gbps links as shown in Figure~\ref{fig:CL1}. In this paper, all the BER measurements are done for 3x$10^{12}$ bits to achieve 95 percent CL. 
Variation of BER at Intel-default transceiver set is recorded with
respect to 
the attenuation of the received optical power; following the methodology flowchart shown in  Figure~\ref{fig:flowchart}. This test is executed with the help of VOA attached to the loopback fibre. 
BER variation is recorded for different PRBS patterns and plotted for
the links operating at 10 Gbps, 4.8 Gbps and 9.6 Gbps rates as shown
in Figure~\ref{fig:bervsdbm_all}.  

The exponential curve fitting is the best-suited approximation for the BER in logarithmic domain~\cite{ippolitoappendix}.
Double exponent fit function with constants is used to fit the BER data as it provides close fits in a variety of BER plot situations. It fits the BER data using unconstrained nonlinear optimization~\cite{NTbook}.
The statistics for goodness-of-fit in terms of R-Square ($R^2$) for
different PRBS is marked in the 
Figure~\ref{fig:bervsdbm_all}. 

The test shown in Figure~\ref{fig:bervsdbm_all} highlights that at a specified CL higher number of errors are received in the transmission system for a given received optical power; when PRBS31 is injected as the test data pattern as compared to the other PRBS patterns.
The outcome of the tests shown in Figure~\ref{fig:prbsall} and Figure~\ref{fig:bervsdbm_all} revealed the degradation of the metrics of signal integrity with the increase in the size of a unique word of data in the PRBS sequence. The results from these tests are as anticipated and well substantiated. It has further strengthened the usefulness of the PRBS31 as a strenuous test pattern to demonstrate the validation of the proposed methodology. However, there is a crossover point for 4.8 Gbps at BER$\sim$$10^{-10}$. It is kept beyond the discussion as our region of interest is better by two orders of magnitude which is BER$\sim$$10^{-12}$.

\subsection{Improvement in Transmission}

\begin{figure}[ht!]
	\begin{center}
		\includegraphics[scale=0.412]{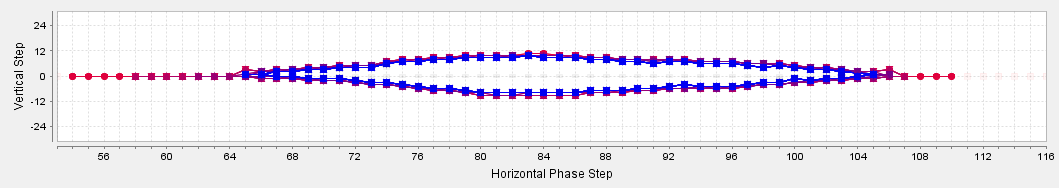}\\
		Vertical step(19)/Horizontal Phase step(41) for 10 Gbps at the Intel FPGA default settings\\
		\includegraphics[scale=0.412]{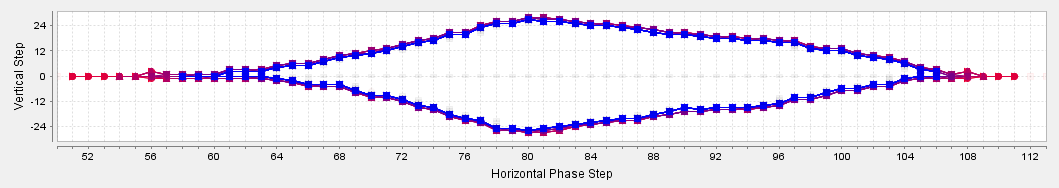}\\
		Vertical step(49)/Horizontal Phase step(54) for 10 Gbps at the Optimized FPGA settings\\
		\includegraphics[scale=0.305]{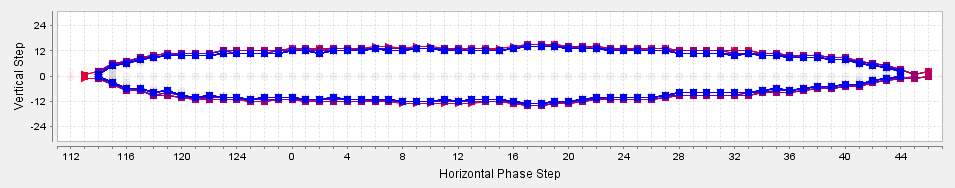}\\
		Vertical step(28)/Horizontal Phase step(59) for 4.8 Gbps at the Intel FPGA default settings\\
		\includegraphics[scale=0.31]{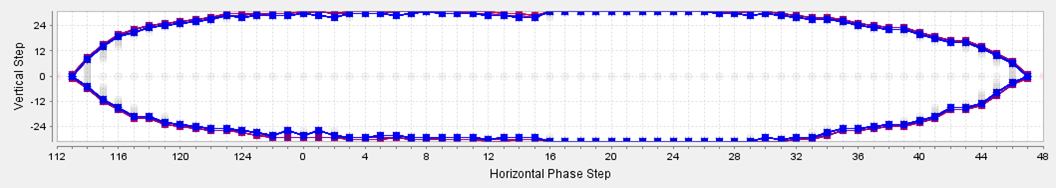}\\
		Vertical step(63)/Horizontal Phase step(63) for 4.8 Gbps at the Optimized FPGA settings\\
		\includegraphics[scale=0.41]{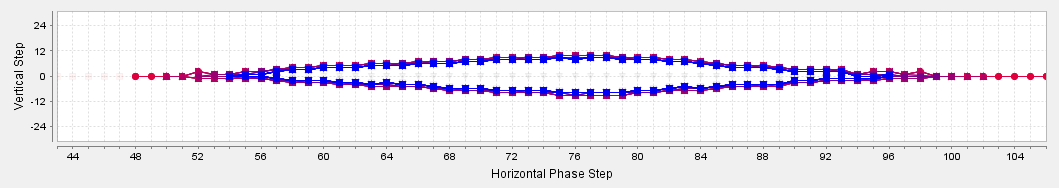}\\
		Vertical step(18)/Horizontal Phase step(43) for 9.6 Gbps at the Intel FPGA default settings\\
		\includegraphics[scale=0.41]{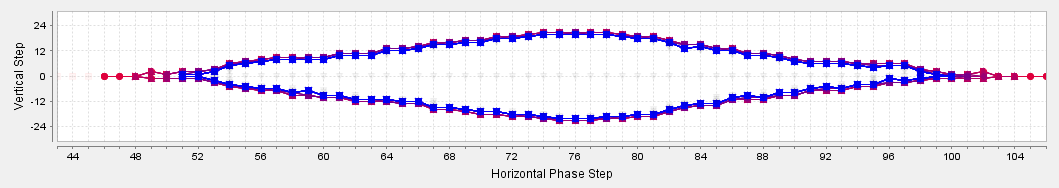}\\
		Vertical step(41)/Horizontal Phase step(50) for 9.6 Gbps at the Optimized FPGA settings\\
		\caption{Eye diagram at the Intel FPGA default and at the Optimized settings of transceiver.}
		\label{fig:eye}
	\end{center}
\end{figure}

The improvement in the system performance is marked by two metrics of signal integrity viz. BER and Eye Diagram. 
The eye contour for the Intel-default settings and at the deduced optimized settings of the transceiver is captured using the EyeQ (a GUI feature of TTK). It helps to estimate and visualize the vertical and horizontal eye opening at the receiver as shown in Figure~\ref{fig:eye}. After the application of the deduced transceiver parameter’s settings using the proposed technique, there is a notable enhancement in width (Horizontal Phase Step) and height (Vertical Step) of the eye diagram. Hence the quality of signal transmission is improved. % and the comparison is clearly marked in the Figure~\ref{fig:threeaxes} using a three axes spider plot with PRBS on the third axis.
%three axis plot

\begin{figure}[htbp]
	\begin{center}
		\includegraphics[trim=3.8cm 1.6cm 3.1cm 1.35cm, clip=true, totalheight=0.29\textheight]{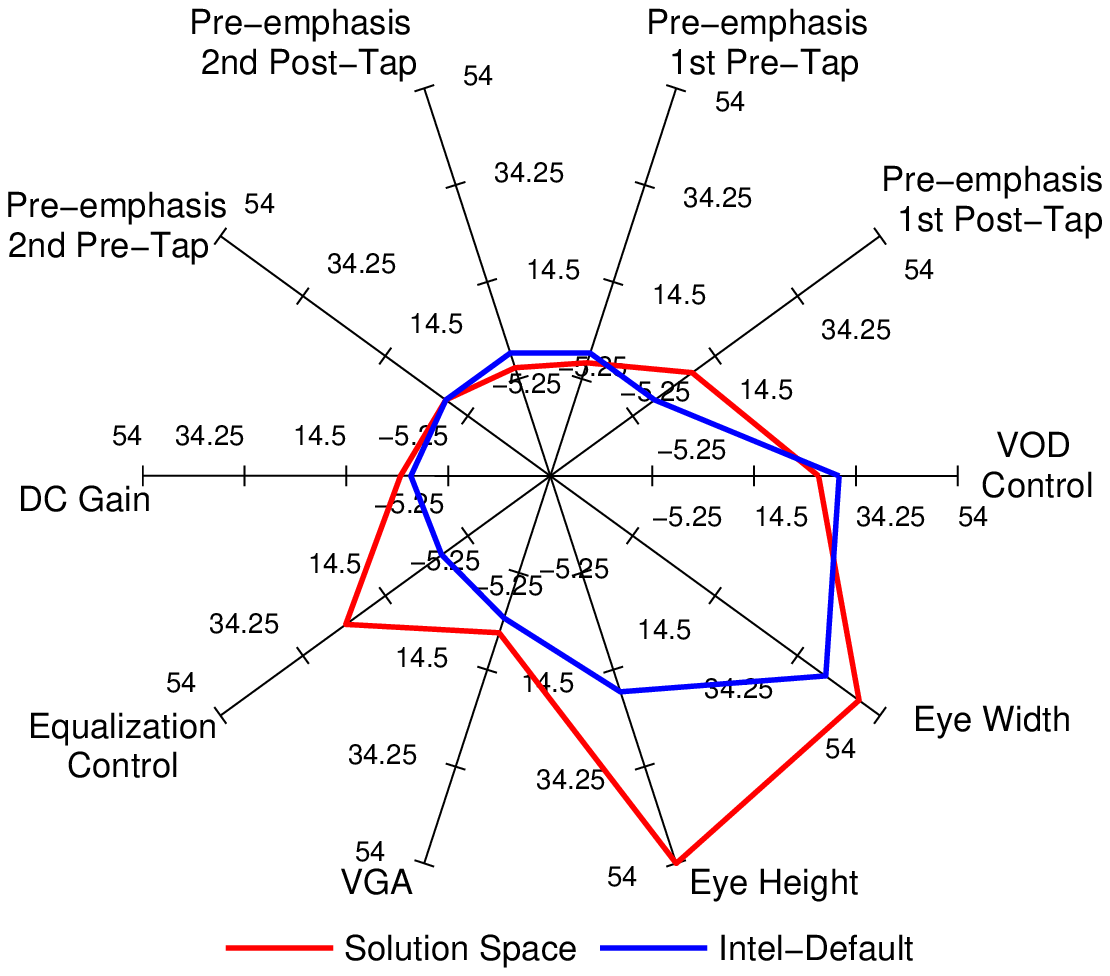}\\
		10 Gbps line rate\\
		\includegraphics[trim=6.6cm 1.6cm 5.9cm 1.35cm, clip=true, totalheight=0.29\textheight]{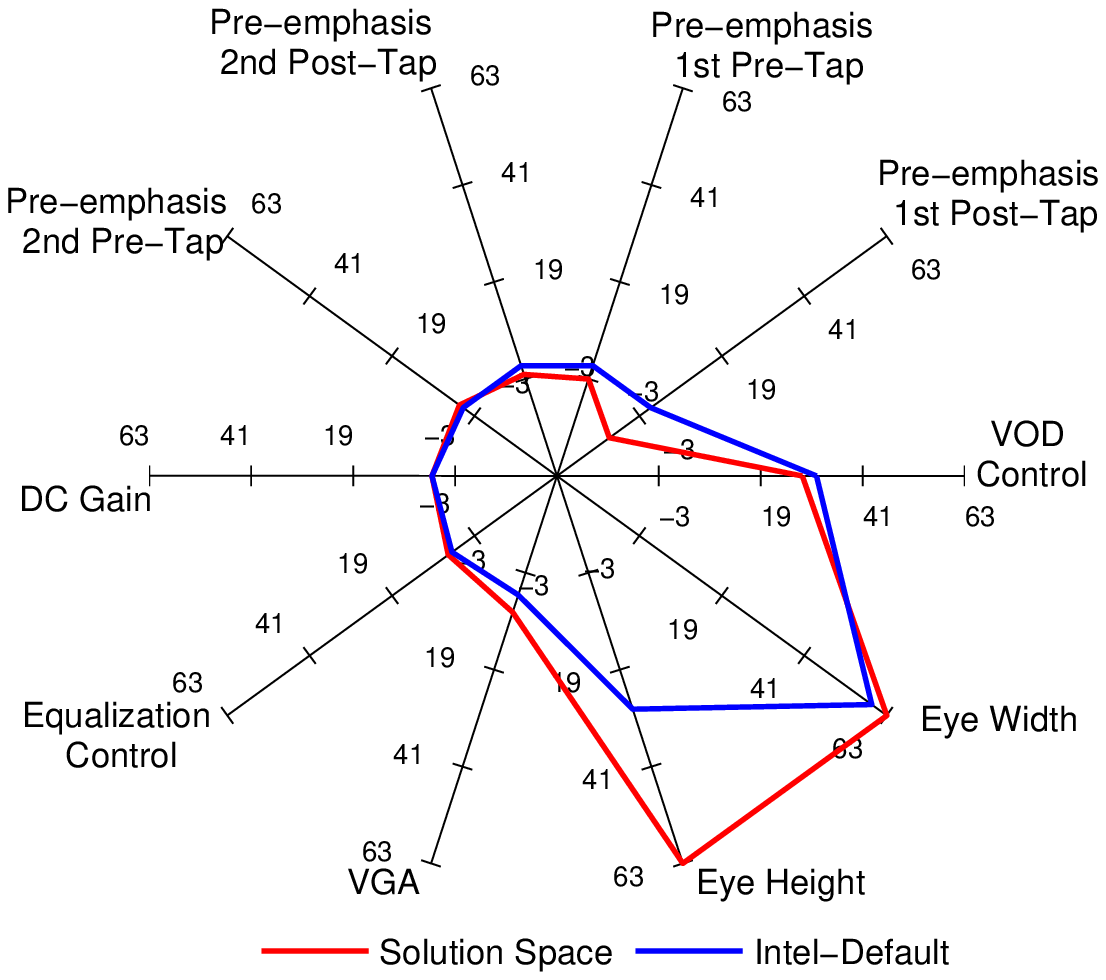}\\
		GBT 4.8 Gbps line rate\\
		\includegraphics[trim=6.00cm 1.6cm 5.40cm 1.35cm, clip=true, totalheight=0.29\textheight]{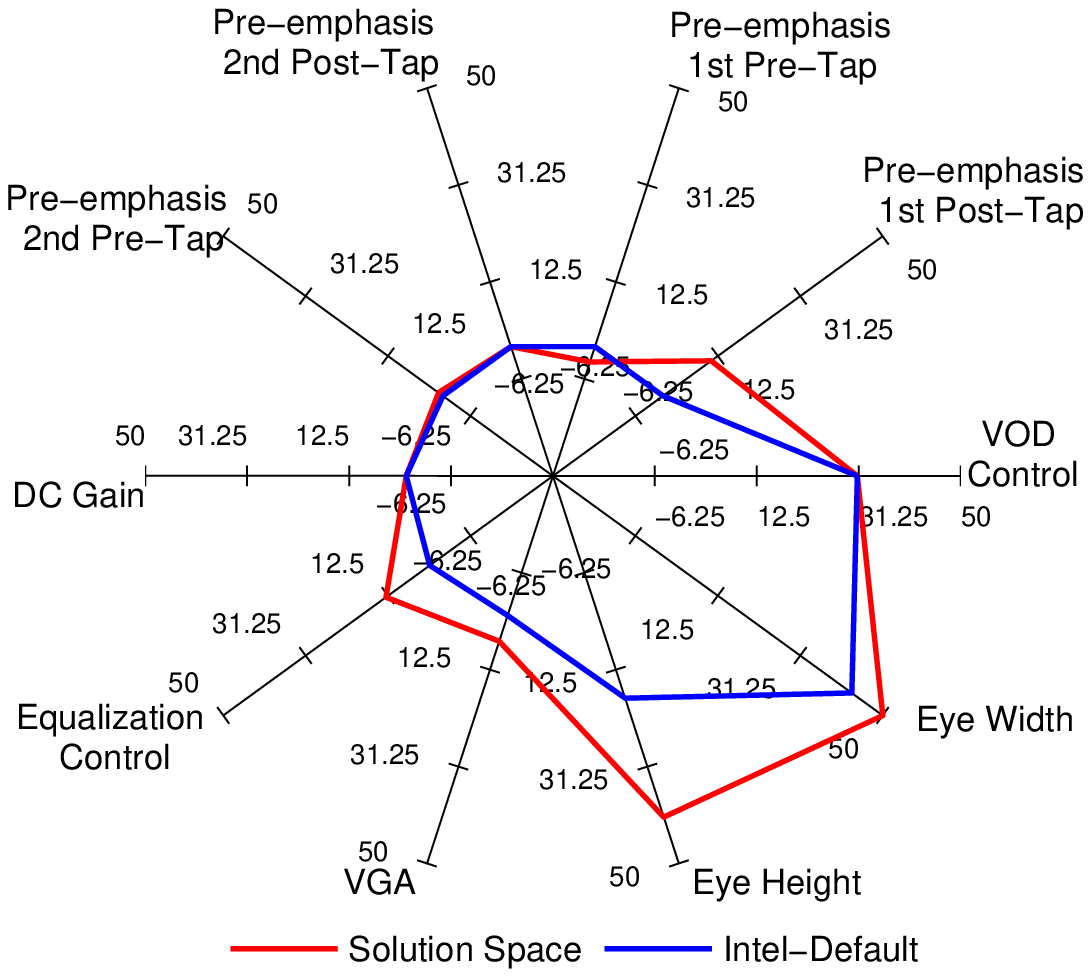}\\
		TTC-PON 9.6 Gbps line rate\\
		\caption{Multivariate kiviat diagram showing the  solution space and the Intel FPGA default values for three different link rates.}
		\label{fig:spider}
	\end{center}
\end{figure}

%\begin{figure}[ht!]
%	\begin{center}
%		\includegraphics[trim=6.6cm 2.6cm 4.4cm 1.83cm, clip=true, totalheight=0.22\textheight]{prbs31_def_opt10g.eps}\\
%		10 Gbps line rate\\
%		\includegraphics[trim=6.6cm 2.6cm 4.4cm 1.83cm, clip=true, totalheight=0.22\textheight]{prbs31_def_opt_gbt.eps}\\
%		GBT 4.8 Gbps line rate\\
%		\includegraphics[trim=7.5cm 2.6cm 5.3cm 1.83cm, clip=true, totalheight=0.22\textheight]{prbs31_def_opt_TTC_1.eps}\\
%		TTC-PON 9.6 Gbps line rate\\
%		\caption{Spider plot for parameter comparison of Eye Diagram at PRBS31 for Default and the Optimized transceiver parameters settings.}
%		\label{fig:threeaxes}
%	\end{center}
%\end{figure}
The optimized values of the transceiver parameters known as solution space, found from the proposed methodology for the targeted BER of $10^{-12}$ are plotted against the Intel-default set in the form of a multivariate kiviat diagram for all the three link speeds as given in Figure~\ref{fig:spider}. It allows us to demonstrate a clear comparison of the individual parameters on each axis.

\begin{figure}[ht!]
	\begin{center}
		\includegraphics[trim=1.8cm 0.2cm 2.45cm 0.12cm, clip=true,scale=0.3]{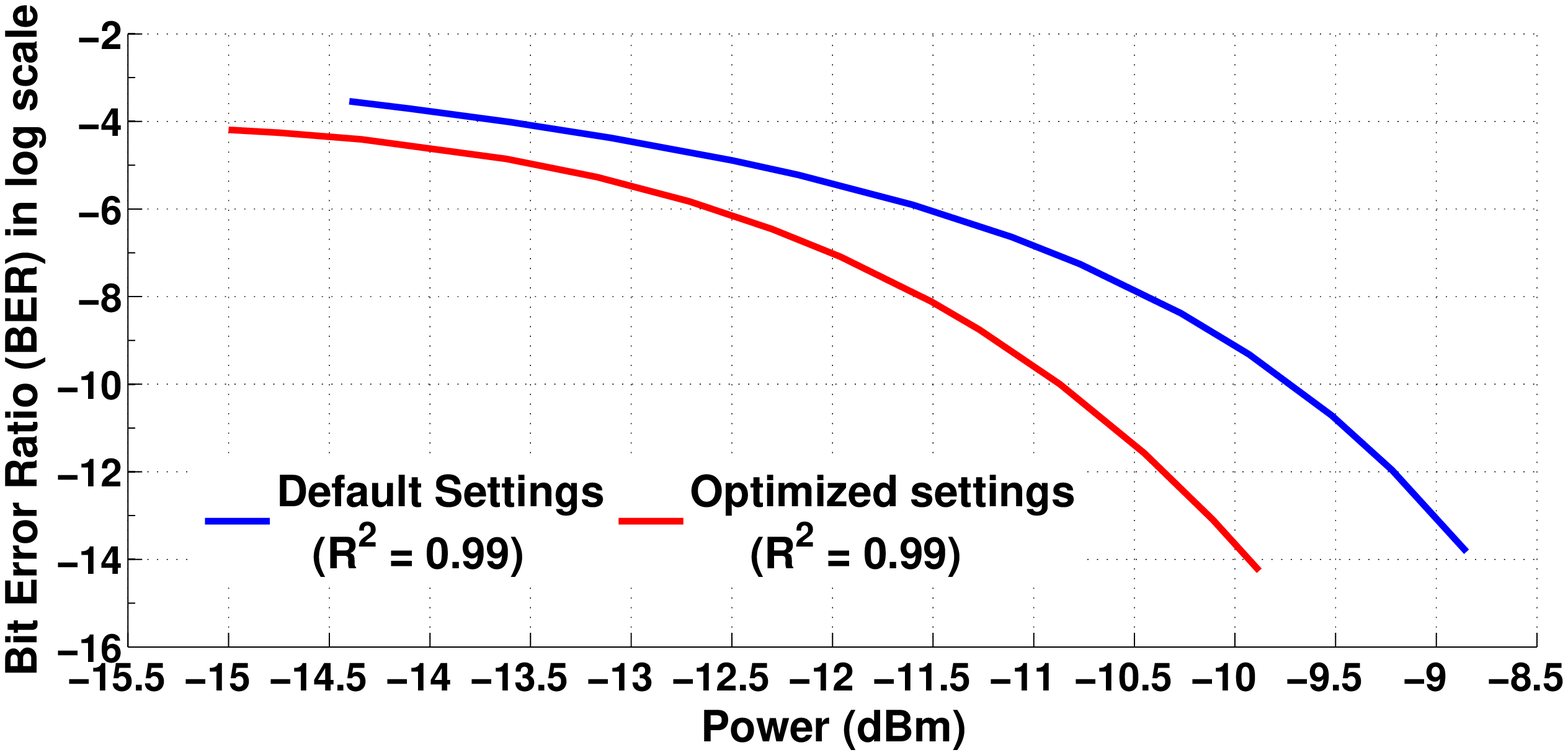}\\
		10 Gbps line rate\\
		\includegraphics[trim=1.95cm 0.2cm 2.6cm 0.05cm, clip=true,scale=0.3]{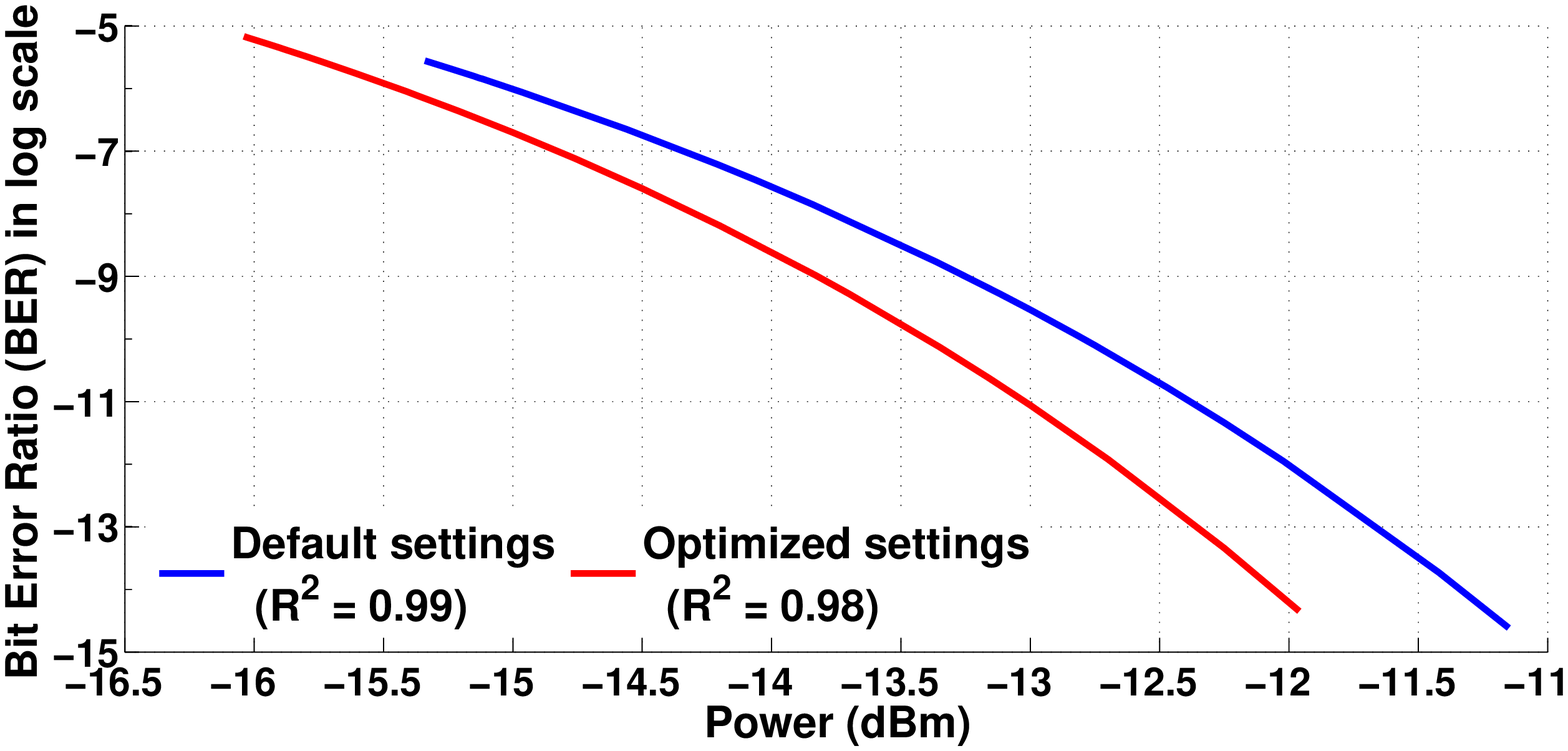}\\
		GBT 4.8 Gbps line rate\\	
		\includegraphics[trim=1.7cm 0.2cm 1.25cm 0.15cm, clip=true,scale=0.29]{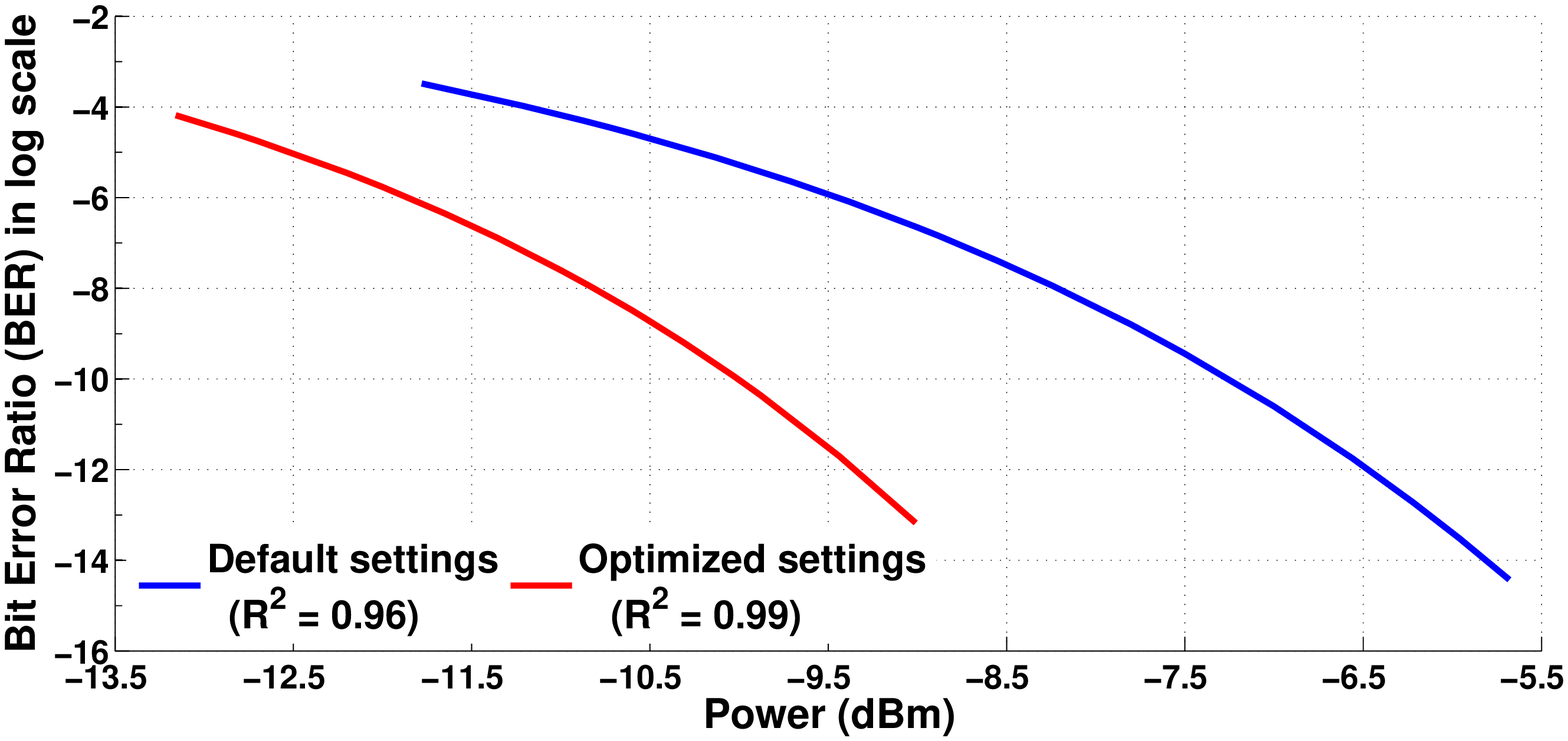}\\
		TTC-PON 9.6 Gbps line rate\\
		\caption{Comparison of BER versus the received optical power for default and optimized transceiver settings separately for three line rates.}
		\label{fig:ber_dbm_opt_def}
	\end{center}	
\end{figure}

Variation in BER is plotted for the deduced solution space values of a transceiver and for the Intel default set; concerning the different attenuation levels of input optical power at the receiver. It is shown for PRBS31 for all the three links under observation in Figure~\ref{fig:ber_dbm_opt_def}.
%bervsdbm prbs31

Further analysing the results from Figure~\ref{fig:ber_dbm_opt_def},
the least optical power required at the receiver to attain a preferred
BER or better could be determined from the curve. Also it shows, that
a specific marked BER is achieved at a lower optical power when
transceiver is operated at the deduced parameter values listed in
solution space in comparison to the Intel default set. Here to
mention the particular case as an example, the targeted BER of
$10^{-12}$ for the optical link test as per IEEE standards is achieved
at lower values of the optical power and the improvement at the
mentioned BER is 
quantitatively listed in Table~\ref{table:opticalpower} for the three link speeds. 

\begin{table}[ht!]
	\centering 
	\caption{Comparison of Optical power(dBm) to attain BER of $10^{-12}$ for the  three high speed interface links.}
	\scalebox{0.69}{
	\begin{tabular}{@{}lllll@{}}
		\toprule
		\begin{tabular}[c]{@{}l@{}}Protocol\end{tabular} & \begin{tabular}[c]{@{}l@{}}With default \\ approach \\(dBm)\end{tabular} & \begin{tabular}[c]{@{}l@{}}With optimization\\ technique \\(dBm)\end{tabular} & \begin{tabular}[c]{@{}l@{}}Difference\\ (dBm)\end{tabular} & \begin{tabular}[c]{@{}l@{}}Improvement \\ (Percentage)\end{tabular} \\ \midrule 
		10Gb Ethernet                                                         & -9.2                                                                   & -10.35                                                                    & -1.15 & 
		{12.5}                                                                \\ \hline
		GBT                                                        & -11.9                                                                  & -12.7                                                                &   -0.8   & 
		{6.7}                                                                 \\ \hline
		TTC-PON                                                       & -6.45                                                                  & -9.3                                                                   &   -2.85  & {44.1}                                                                \\ \bottomrule
	\end{tabular}}
	\label{table:opticalpower}
\end{table}

Another clear observation emerged from the data comparison of
Figure~\ref{fig:ber_dbm_opt_def} is that 
the receiver sensitivity below which the loss of lock occurs, is
enhanced due to the reduction in the high-frequency losses with the
application of the proposed optimization technique. This results in
reducing the limit of the optical power required for the proper CDR
and the signal is traceable for comparatively lower values of the
received optical power. The quantitative comparisons are given in
Table~\ref{table:Receiver_Sensitivity}.

\begin{table}[ht!]
	\centering 
	\caption{Comparison of optical power for CDR for the  three high speed interface links.}
		\scalebox{0.69}
{	\begin{tabular}{@{}lllll@{}}
		\toprule 
		{\begin{tabular}[c]{@{}l@{}}Protocol\end{tabular}} & {\begin{tabular}[c]{@{}l@{}}With default \\ parameters\\ (dBm)\end{tabular}} & {\begin{tabular}[c]{@{}l@{}}With optimization\\ technique \\(dBm)\end{tabular}} & \begin{tabular}[c]{@{}l@{}}Difference\\ (dBm)\end{tabular} & \begin{tabular}[c]{@{}l@{}}Improvement\\ (Percentage)\end{tabular} \\ \midrule 
		10Gb Ethernet                                                                  & -14.4                                                                             & -15                                                                                  & -0.6 & {4.17}                                                               \\  \hline
		GBT                                                                 & -15.34                                                                            & -16.04                                                                               & -0.7 & {4.56}                                                               \\  \hline
		TTC-PON                                                           & -11.78                                                                            & -13.2                                                                                & -1.42 & {12.05}                                                              \\ \bottomrule 
	\end{tabular} 
\label{table:Receiver_Sensitivity}}
\end{table}

The test results shown in Figure~\ref{fig:spider} and~\ref{fig:ber_dbm_opt_def}  
confirms that the effect of high-frequency losses on the link
performance is controlled. It is achieved after the application of the
deduced solution space values to the TTK and a significant improvement
on the BER is noted at a particular received optical power.
The tests and results validate the usefulness of the proposed
technique to enhance the transceiver performance and the signal
integrity by compensating for the high-frequency losses.

\section{Summary}\label{summary}

We have presented a novel transceiver optimization technique to 
reduce the high-frequency losses which occur due to the increased rates 
of data transmission in case of HEP experiments.
The technique has been implemented on the latest 20nm Intel-Altera
Arria-10 FPGA. The scheme has been tested and validated for the link
rates of three high-speed communication protocols,  GBT, TTC-PON and
10 Gbps Ethernet, which are most commonly used for interfacing the
detector front-end electronics, trigger and DAQ systems. The proposed
scheme is an optimized approach which reduces 
number of iterations required.

The tests are performed with PRBS31 pattern at a confidence level of
95 percent. There is considerable gain in the system performance
with the application of the proposed technique as specified by the two
parameters of signal integrity, the BER and the Eye Diagram. The Intel
FPGA set parameters and the solution space values are marked on the
kiviat diagram for the fast comparison between the parameters. The
results point that to attain the marked BER of $10^{-12}$; the
required optical power is reduced by 12.5\%, 6.7\% and 44.1\% for
10Gbps, GBT and TTC-PON respectively. The BER is also improved over
the received range of optical power. The CDR capability of the system
is also enhanced as the least optical power required to recover the
data traffic is reduced by 4.17\%, 4.56\% and 12.05\% for 10Gbps,
GBT and TTC-PON respectively. The technique improves the signal
integrity and reduces the BER. 
This technique is a heuristic solution and has potential for practical applications as it provides rapid convergence of the solution space to achieve optimized transceiver settings.
%This technique is a heuristic solution for rapid optimization of the transceivers. 
It makes the
implementation of the new technique time
efficient. This transceiver optimization technique and its
implementation approach would lend itself 
well for other FPGAs users that allows on-chip assessment of signal quality like Eye diagram.

{\bf Acknowledgement}

The authors gratefully acknowledge the support of the ALICE
Collaboration at CERN during the period of the research work. We thank Alex
Kluge, Tivadar Kiss, Erno David of the ALICE Electronics coordination
and the CRU project for their valuable help and advice. We thank Subhasis Chattopadhyay,
Anurag Misra and Saurabh Srivastava for fruitful suggestions during
the preparation of the manuscript.

\bibliography{reference}	
\bibliographystyle{elsarticle-num}
\end{document}